\documentclass[a4paper,11pt]{article}
\usepackage{jheppub}
\usepackage{mathrsfs}
\usepackage{amsfonts}
\usepackage{setspace}
\usepackage{cellspace}
\usepackage{amsmath,amssymb,bm}
\usepackage[colorlinks=true,linkcolor=blue]{hyperref}
\usepackage{xcolor}
\usepackage{epsfig}
\usepackage{slashed}
\usepackage{caption}
\usepackage{hhline,multirow,tabularx}  
\usepackage{dcolumn}    
\usepackage{url}        
\usepackage{braket} 
\definecolor{cyan}{rgb}{0.0, 1.0, 1.0}
\definecolor{applegreen}{rgb}{0.55, 0.71, 0.0}
\definecolor{arylideyellow}{rgb}{0.91, 0.84, 0.42}
\definecolor{bananayellow}{rgb}{1.0, 0.88, 0.21}
\definecolor{burlywood}{rgb}{0.87, 0.72, 0.53}
\definecolor{buff}{rgb}{0.94, 0.86, 0.51}
\definecolor{blond}{rgb}{0.98, 0.94, 0.75}
\definecolor{bisque}{rgb}{1.0, 0.89, 0.77}
\definecolor{bananamania}{rgb}{0.98, 0.91, 0.71}
\definecolor{apricot}{rgb}{0.98, 0.81, 0.69}
\definecolor{almond}{rgb}{0.94, 0.87, 0.8}

\title{Threshold resummation for computing large-$x$ parton distribution through large-momentum effective theory}
\author[a]{Xiangdong Ji}
\author[b]{Yizhuang Liu}
\author[a]{Yushan Su}

\affiliation[a]{Department of Physics, University of Maryland, College Park, MD 20742}
\affiliation[b]{Institute of Theoretical Physics,
Jagiellonian University, 30-348 Kraków, Poland}

\emailAdd{xji@umd.edu}
\emailAdd{yizhuang.liu@uj.edu.pl}
\emailAdd{ysu12345@umd.edu}

\abstract{Parton distribution functions (PDFs) at large $x$
are poorly constrained by high-energy experimental data, but extremely important for probing physics beyond standard model at colliders. We study the calculation of PDFs at large-$x$ through large-momentum $P^z$ expansion of the lattice quasi PDFs. Similar to deep-inelastic scattering, there are two distinct perturbative scales in the threshold limit where the matching coefficient can be factorized into a space-like jet function at scale $P^z|1-y|$ and a pair of heavy-light Sudakov form factors at scale $P^z$. 
The matching formula allows us to derive a full renormalization group resummation of large threshold logarithms, and the result is consistent with the known calculation to the next-to-next to leading order (NNLO). This 
paves the way for direct large-$x$
PDFs calculations in lattice QCD. As by-products, we find that the space-like jet function is related to a time-like version calculated previously through analytic continuation, and the heavy-light Sudakov form factor, calculated here to NNLO, is a universal object appearing as well in the large momentum expansion of quasi transverse-momentum-dependent PDFs and quasi wave-function amplitudes.}
\date{\today}

\begin{document}
\maketitle
\flushbottom
\section{Introduction}
The parton distribution functions (PDFs) are universal objects in high energy quantum chromodynamics (QCD). One one hand, they are non-perturbative scaling functions underlining the prior-QCD Bjorken-scaling relation~\cite{Bjorken:1968dy} for total cross section of $ep$ collision. One the other hand, in modern languages they are intrinsic non-perturbative distribution functions describing the internal distribution of quark and gluons in hadrons (e.g, proton). In the large $Q^2$ limit of many experimental processes ranging from inclusive deep-inelastic scattering (DIS) to the Drell-Yan (DY) pair production, the total cross sections can be expressed in terms of perturbative $Q^2$ dependent hard coefficients convoluted with the PDFs, allowing the later to be extracted from experimental data~\cite{Dulat:2015mca, Hou:2019efy, Accardi:2016qay, H1:2015ubc, Alekhin:2017kpj, Jimenez-Delgado:2014twa,Harland-Lang:2014zoa,NNPDF:2017mvq}. 

However, the extraction of PDF at large-$x$ still suffers from large uncertainties (e.g. 5$\%$-10$\%$ relative uncertainties for $x \gtrsim 0.7$),  although this region of $x$ is important for understanding property of strong interaction such as color confinement~\cite{Holt:2010vj}, isospin dependence~\cite{Afnan:2003vh,Tropiano:2018quk,JeffersonLabHallATritium:2021usd}, spin structure~\cite{JeffersonLabHallA:2016neg,STAR:2019yqm,Friscic:2021oti,Lagerquist:2022tml} and EMC effect~\cite{EuropeanMuon:1983wih,Ke:2023xeo}, as well as to detect beyond standard model effects in colliders~\cite{Kuhlmann:1999sf,CMS:2012ftr,ATLAS:2011juz}. One particular reason for the lacking of accuracy for large-$x$ PDF extraction is due to additional complexity on the theoretical side: in addition to the standard large $\frac{Q^2}{\Lambda_{\rm QCD}^2}$ logarithms that can be controlled by OPE and asymptotic freedom, near threshold region for various processes such as DIS, DY and Higgs boson production, there are threshold logarithms $\ln^n (1-x)$, due to presence of  additional scales of the form $(1-x)Q^2$, and requires re-summation in order to make controlled prediction.  The attempt trying to re-sum these threshold logarithms dates back to early days of QCD~\cite{Sterman:1986aj,Catani:1989ne} and has been investigated more systematically~\cite{Bauer:2002nz,Manohar:2003vb,Pecjak:2005uh,Chay:2005rz,Idilbi:2006dg,Chen:2006vd, Becher:2006mr,Bonvini:2012az} in the framework of soft-collinear effective theory (SCET) and renormalization group equations~\cite{Bauer:2000yr,Bauer:2001yt,Becher:2014oda}. At practical level,  threshold resummation has been implemented in some global fittings such as~\cite{Corcella:2005us,Aicher:2010cb,Bonvini:2015ira,Westmark:2017uig,Barry:2021osv}, but uncertainties are still large.

In recent years, large-momentum effective theory (LaMET)~\cite{Ji:2013dva,Ji:2014gla,Cichy:2018mum,Ji:2020ect} has been proposed to provide an effective method to calculate the $x$-dependence of PDFs directly from lattice QCD without fitting of unknown functional forms. For example, quark isovector PDFs have been calculated based on LaMET~\cite{Lin:2014zya,Alexandrou:2015rja,Chen:2016utp,Alexandrou:2016jqi,Alexandrou:2018pbm,Chen:2018xof,Lin:2018pvv,LatticeParton:2018gjr,Alexandrou:2018eet,Liu:2018hxv,Chen:2018fwa,Izubuchi:2019lyk,Shugert:2020tgq,Chai:2020nxw,Lin:2020ssv,Fan:2020nzz,Gao:2021dbh,Gao:2022iex,Su:2022fiu,LatticeParton:2022xsd,Gao:2022uhg}. The most common choice in LaMET applications is to start from the equal-time matrix element motivated from ordinary momentum distribution (quasi-PDF)
\begin{align}\label{eq:defquasi}
\tilde f \left(y,\frac{P^z}{\mu}\right)=P^z\int_{-\infty}^{\infty} \frac{dz}{2\pi}e^{iyP^z z}\bigg\langle P \bigg|\bar \psi(z)\gamma^t{\cal P}\exp \bigg[-ig\int_{0}^z A^z(z') dz' \bigg]\psi(0) \bigg|P\bigg\rangle \ ,
\end{align}
describing the distribution of $k^z$ of quark-parton inside the hadron state $| P\rangle$. At large $P^z$, the large momentum expansion allows the quasi-PDF to be expressed in terms of the non-perturbative PDFs and perturbative matching coefficients, in a way similar to that of the DIS structure function
\begin{align}\label{eq:factoP}
\tilde f\left(y,\frac{P^z}{\mu}\right)=\int_{-1}^{1} \frac{dx}{|x|}{\cal C}\bigg(\frac{y}{x},\frac{xP^z}{\mu}\bigg)f(x,\mu)+{\cal O}\bigg(\frac{\Lambda_{QCD}^2}{P_z^2}\bigg) \ .
\end{align}
However, it is known~\cite{Xiong:2013bka,Gao:2021hxl} that in the matching kernel ${\cal C}\left(\xi,\frac{xP^z}{\mu} \right)$, to arbitrary order in perturbation theory (PT) there are threshold singularities as $\xi \rightarrow 1$ of the form $\frac{\ln^n |1-\xi|}{1-\xi}$, sometimes called threshold logarithms. These singularities blow up as the order of perturbation theory increases and must be re-summed for precise calculations of quark-PDF $f(x)$ even at moderate $x$. 

The main goal of this paper is to perform a systematic study/resummation of these threshold logarithms in the perturbative matching kernel ${\cal C}$ in case of quark non-singlet PDF. It is sufficient to study the $y\rightarrow 1$ limit for the perturbative quark quasi-PDF in an on-shell quark state with momentum $p^z>0$, and then recover ${\cal C}$ by the substitution $p^z \equiv |xP^z|$ and $y\rightarrow \frac{y}{x}=\xi$. It turns out that all the contributions at power $(1-\xi)^{-1}$, including all the threshold logarithms as well as the $\delta(1-\xi)$ term can be re-summed through the factorization formula
\begin{align}\label{eq:threshPT}
{\cal C}\left(\xi,\frac{p^z}{\mu}\right)\bigg|_{\xi \rightarrow 1}=H\left(\frac{4p_z^2}{\mu^2}\right)p^zJ_f\left(\frac{(1-\xi)p^z}{\mu},\frac{4p_z^2}{\mu^2}\right)\ ,
\end{align}
to all orders in perturbation theory. 
\begin{figure}[h!]
    \centering
    \includegraphics[height=6cm]{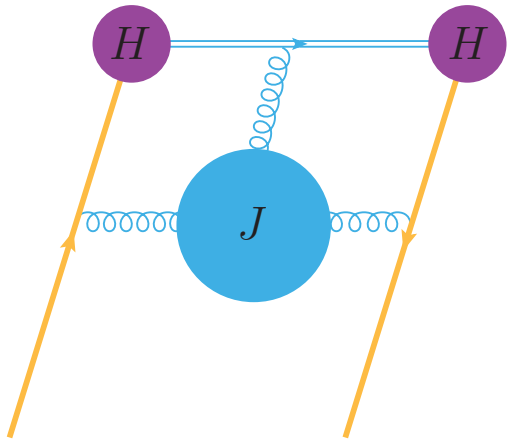}
    \caption{The depiction of the threshold factorization of the quark quasi-PDF. The orange lines represent the incoming and outgoing collinear quark, the purple blobs represent the hard kernel $H$, and the jet function, formed by gluons in the blue blob re-sums all the semi-hard exchanges at scale $|1-\xi|p^z$ between the collinear quark and the static gauge link.}
    \label{fig:threshold}
\end{figure}
The above formula is based on the leading region of the threshold limit for quasi-PDF as shown in Fig.~\ref{fig:threshold} and is similar in spirit to that for deep inelastic scattering (DIS)~\cite{Sterman:1986aj,Becher:2006mr}.  The hard kernel $H$ is real, and is the absolute value of a pair of universal heavy-light Sudakov form factors, due to the hard exchanges around the quark-link vertices at hard scale $4p^z$. On the other hand, the jet function $J_f$ (defined in Eq.~(\ref{eq:defJf})) combines a bare jet function $\tilde J$ (defined in Eq.~(\ref{eq:deftildeJ})) re-summing all the soft gluon exchanges at the semi-hard or intermediate scale $|\xi-1|p^z$, together with the phase of the heavy-light Sudakov form factors.  In the rest of the paper we will present their definitions,  renormalization group equations (RGEs) and anomalous dimensions, and make explicit comparison with the exact NNLO result of quasi-PDF~\cite{Li:2020xml,Chen:2020ody}.

The organization of the paper is as follows. In Sec.~\ref{sec:summary}, we summarize our main results including the factorization formula in the threshold limit as well as RGE resummed form of the matching kernel. 

In Sec.~\ref{sec:space-like jet}, we discuss the spacetime picture underling the threshold limit of the DIS and the quasi-PDF, emphasizing the emergence of scale-separation. This naturally leads us to define a space-like jet function $\tilde J(\mu z)$ in terms of light-like and space-like gauge-links. We present its result at NLO, and show that it relates to time-like jet function in Ref.~\cite{Jain:2008gb} by a simple analytic continuation at NNLO.  We also present its RGE and anomalous dimensions. 

In Sec.~\ref{sec:suda}, we review the heavy-light Sudakov form factor that appears in large momentum expansion of quasi-TMDPDF~\cite{Ji:2014hxa,Ebert:2019okf,Ji:2019ewn,Vladimirov:2020ofp,Ebert:2020gxr,Ebert:2022fmh} and quasi-LFWF~\cite{Ji:2021znw}, in particular, its imaginary part. We redefine the jet function by absorbing the phase of the hard kernel into jet function, leaving the hard kernel real and the same as the hard kernel for quasi-TMDPDF. We show that our NLO jet function and hard kernel reproduces the threshold limit of quasi-PDF. 

In Sec.~\ref{sec:NNLO}, we investigate the threshold limit at NNLO, utilizing the known two-loop anomalous dimensions~\cite{Ji:2019ewn,Ji:2021znw,LPC:2022zci} and the NNLO jet function in Ref.~\cite{Jain:2008gb}. We show that our prediction completely agrees with the threshold limit extracted directly form the full two-loop matching kernel~\cite{Li:2020xml,Chen:2020ody}. Furthermore, we extracted the constant terms of the NNLO heavy-light Sudakov form factor. 

In Sec.~\ref{sec: evo}, we investigate further the RGEs and anomalous dimensions. Using relations among the anomalous dimensions, we show that the RGE of the perturbative matching kernel in the threshold limit matches precisely with the threshold limit of the DGLAP kernel. Using the RGE of the hard kernel and jet function, we derive the fully resummed form of the matching kernel in momentum space. 

Finally, we conclude in Sec.~\ref{sec:conc}. Some technical details are collected in Appendices.

\section{Summary of the main results}\label{sec:summary}

In this section we summarize the main results of the paper, without mentioning $\alpha\equiv \alpha_s=\frac{g^2}{4\pi}$ is the strong-coupling constant.
At the heart of the paper is the factorization formula, stating that the $\overline{\rm MS}$ scheme matching kernel between quasi-PDF and PDF, defined in Eq.~(\ref{eq:factoP}), in the threshold limit $\xi \rightarrow 1$, factorizes into two heavy-light Sudakov form factors $H$ and a space-like jet function $J_f$ as:
\begin{align}\label{eq:finalmatchingC}
{\cal C}\left(\xi,\frac{xP^z}{\mu}\right)\bigg|_{\xi \rightarrow 1}=H\left(\frac{4x^2P_z^2}{\mu^2}\right)|xP^z|J_f\left(\frac{(1-\xi)xP^z}{\mu},\frac{4x^2P_z^2}{\mu^2}\right)  \ .
\end{align}
In this formula, one has
\begin{enumerate}
    \item $H\left(\frac{4xP_z^2}{\mu^2}\right)$ is the heavy-light Sudakov form factor, re-summing the hard exchanges at the scale $\zeta_z=4x^2P_z^2$ around the quark-link vertices. The renormalization group equation for the hard kernel reads 
\begin{align}
\frac{d}{d\ln \mu} \ln H\left(\frac{\zeta_z}{\mu^2},\alpha_s(\mu)\right)=\Gamma_{\rm cusp}(\alpha_s)\ln\frac{\zeta_z}{\mu^2}+\tilde {\gamma}_{H}(\alpha_s) \ ,
\end{align}    
where $\Gamma_{\rm cusp}$ is the light-like cusp anomalous dimension~\cite{Korchemsky:1987wg} and is known up to four loops~\cite{Henn:2019swt,vonManteuffel:2020vjv}, and $\tilde \gamma_{H}$ is the same anomalous dimension $\gamma_{C}$ arising in the context of large momentum expansion of transverse momentum dependent quantities~\cite{Ji:2019ewn,Ji:2020ect,Ji:2021znw}.
The explicit formulas for $\Gamma_{\rm cusp}$ and $\tilde \gamma_H$ up to two loops are given in Eq.~(\ref{eq:resultcusp}) and Eq.~(\ref{eq:tildegammaHresult}) respectively.
\item The jet function $J_f\left(\frac{(1-\xi)xP^z}{\mu},\frac{4x^2 P_z^2}{\mu^2}\right)$ is another perturbative calculable object.   It combines a space-like jet function (defined in Eq.~(\ref{eq:deftildeJ})) re-summing all the soft gluon exchanges at the semi-hard scale $\mu_i=|\xi-1||xP^z|$ between the incoming collinear quark and the static gauge-link, together with the phase of the heavy-light Sudakov form factors at the scale $\zeta_z$.  It is perturbative calculable when $\Lambda_{\rm QCD}\ll |\xi-1||xP^z| \ll |xP^z|$.  The evolution equation for the jet function in momentum space reads
\begin{align}
&\frac{d}{d\ln \mu} J_{f}\left(\frac{(1-\xi)xP^z}{\mu},\frac{4x^2P_z^2}{\mu^2}\right)=-\left(\tilde \gamma_{J}+\Gamma_{\rm cusp}\ln \frac{4x^2 P_z^2}{\mu^2}\right)J_f\left(\frac{(1-\xi)xP^z}{\mu},\frac{4x^2P_z^2}{\mu^2}\right)
\nonumber \\ 
&-2\Gamma_{\rm cusp}{\cal P}\int_{\xi'>\xi} \frac{J_f\left(\frac{(1-\xi')x P^z}{\mu},\frac{4x^2P_z^2}{\mu^2}\right)}{\xi'-\xi}d\xi' \ ,
\end{align}     
where $\tilde \gamma_J$ is another anomalous dimension given explicitly in Eq.~(\ref{eq:resulttildegammaJ}), and our principle value or plus function ${\cal P}$ is defined in 
Eq.~(\ref{eq: defprin}).
\end{enumerate}

Given the RGE of the various pieces, one can resum large logarithms by evolving from the natural scales of each individual pieces to the common renormalization scale $\mu$ that matches with the PDF.  More precisely, in the Sudakov form factor (including the phase factor) one evolves from $\zeta_z$ to $\mu$, while in the jet function $\tilde J$ (without the phase factor) one evolves from $\mu_i=|1-\xi||xP^z|$ to $\mu$. In practical calculations, one can choose $\mu_i=|1-y| P^z$ during matching or $\mu_i=|1-x| P^z$ during the inverse matching to avoid Landau pole singularities, and the differences of scale choice during the convolutions are higher ${\cal O}(1-y)$ corrections in the threshold factorization. We will test the scale choice numerically in a later work. The details of the evolution are similar to that in~\cite{Becher:2006mr} and provided in Sec.~\ref{sec: evo}.  To summarize, the RGE resummed form of the matching coefficient reads
\begin{align}
&{\cal C}\left(\xi,\frac{xP^z}{\mu}\right)=H(\alpha(\zeta_z))\exp \bigg[2S(\zeta_z,\mu)-2S(\mu_i,\mu)-a_{H}(\zeta_z,\mu)+a_{J}(\mu_i,\mu)\bigg] \nonumber \\
&\times \tilde J\left(l_z=-2 \partial_\eta,\alpha(\mu_i)\right) \left[\frac{\sin \left({\rm sign}(1-\xi)\hat A(\zeta_z,\mu)+\frac{\eta \pi}{2}\right)}{|1-\xi|}\left(\frac{2|1-\xi||xP^z|}{\mu_i}\right)^\eta\right]_{*}\frac{ \Gamma(1-\eta) \mathrm{e}^{-\eta \gamma_E}}{\pi} \ .
\end{align}
In the above formula $[ \ ]_{\star}$ is the star operation defined in Eq.~(\ref{eq:star}), and   
\begin{enumerate}
     \item The evolution factors due to $\Gamma_{\rm cusp}$~\cite{Becher:2006mr} read
\begin{align}\label{eq:aGamma}
S(\nu,\mu)=-\int_{\alpha(\nu)}^{\alpha(\mu)}\frac{\Gamma_{\rm cusp}(\alpha)d\alpha}{\beta(\alpha)}\int_{\alpha(\nu)}^{\alpha}\frac{d\alpha'}{\beta(\alpha')} \ , \ a_{\rm \Gamma}(\nu,\mu)=-\int_{\alpha(\nu)}^{\alpha(\mu)} d\alpha\frac{\Gamma_{\rm cusp}(\alpha)}{\beta(\alpha)} \ .
\end{align}
In term of $\mu_i$ one has $\eta=2a_{\Gamma}(\mu_i,\mu)$.
The other single log evolution factors~\cite{Becher:2006mr} reads
\begin{align}\label{eq:othera}
&a_{H}(\nu,\mu)=-\int_{\alpha(\nu)}^{\alpha(\mu)} d\alpha\frac{\tilde \gamma_H(\alpha)}{\beta(\alpha)} \ ,  \ 
a_{J}(\nu,\mu)=-\int_{\alpha(\nu)}^{\alpha(\mu)} d\alpha\frac{\tilde \gamma_{J}(\alpha)}{\beta(\alpha)} \ ,
\end{align}    
where our convention for $\beta$ can be found in Eq.~(\ref{eq:beta}).
\item $H(\alpha(\zeta_z))$ is the real part of the hard kernel evaluated at scale $\zeta_z$, which can be found up to NNLO in Eq.~(\ref{hardoneloop}), Eq.~(\ref{hardtwoloop}) and Eq.~(\ref{eq:resultcH}). $\tilde J(l_z,\alpha(\mu_i))$ is a jet function in coordinate space at scale $\mu_i$ in $\overline{\rm MS}$ scheme, which can be found in Eq.~(\ref{jetoneloop}), Eq.~(\ref{jettwoloop}) and Eq.~(\ref{eq:resultc1}). One can modify it to hybrid scheme in practical calculations. Finally, 
\begin{align}
\hat A(\zeta_z,\mu)=A(\alpha(\zeta_z))+\pi a_{\rm \Gamma}(\zeta_z,\mu) \ ,
\end{align} 
where $A(\alpha(\zeta_z))$ is the phase angle of the imaginary part evaluated at scale $\zeta_z$, which can be found in Eq.~(\ref{eq:resultJf2}), Eq.~(\ref{eq:resultJf1}) and Eq.~(\ref{eq:resultca}).
\end{enumerate}
In a future publication we will apply our threshold resummation formalism to numerical applications of pion and proton PDFs.

\section{Space-time picture of the threshold limit and the space-like heavy-quark jet function}\label{sec:space-like jet}

In this section we study the space-like jet function, a crucial object in order to resum the large threshold logarithm.  Our convention for the light-cone plus vector is $n=\frac{1}{\sqrt{2}}(1,1,0,0)$ and the space-like vector is $n_z=(0,1,0,0)$.  In order to be pedagogical we first provide a brief introduction to the threshold factorization of DIS induced by an incoming quark with four-momenta $p=(p^+,0,0,0)$ and a virtual photon with four-momenta $q=(-x_Bp^+,x_Bp^+,0,0)$. As usual, $-q^2=Q^2=2x_B^2(p^+)^2$ is large and $x_B=-\frac{q^2}{2P\cdot q}$ is the standard Bjorken $x$.
\begin{figure}[h!]
    \centering
    \includegraphics[height=5cm]{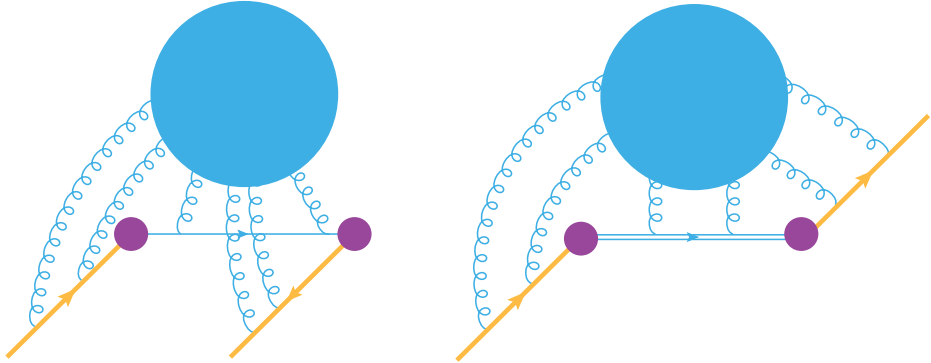}
    \caption{The space-time picture for DIS (left) and quasi-PDF (right) in the threshold limit. The $t$-direction is vertical while the $z$-direction is horizontal. The orange line represents the collinear quarks. The blue single (double) line represents the anti-collinear quark (space-like gauge-link). Blue blobs represent gluon exchanges at scale $\mu_i$ or $\mu_{i,{\rm DIS}}$. Finally, the purple blobs show where Sudakov hard exchanges happen.}
    \label{fig:DISquasi}
\end{figure}

Unlike the collinear factorization of DIS at a generic $x_B$, in the threshold $x_B \sim 1$ region~\cite{Sterman:1986aj,Catani:1989ne,Bauer:2002nz,Manohar:2003vb,Pecjak:2005uh,Chay:2005rz,Idilbi:2006dg,Chen:2006vd, Becher:2006mr,Bonvini:2012az}, all the collinear splittings $q\rightarrow q+g$ along the trajectory of the incoming quark have very small gluon momentum fraction, or $p_g^+=zp^+$ with $z \sim x_B \ll 1$. Due to this, the incoming quark almost loses no collinear momenta, until hit by the virtual photon and turns into an almost anti-collinear one with $p+q\sim(0,1,0,0)p^+$, inducing the final state jet. This leads to strong Sudakov effect~\cite{Sudakov:1954sw, Collins:1989bt} caused by hard exchanges that are purely virtual, at the scale $Q^2$.  
On the other hand, the final state jet, consisting of the anti-collinear quark and all the real ``soft gluon emmisions'' has a new perturbative scale
\begin{align}
\Lambda_{\rm QCD}^2\ll \mu_{i, {\rm DIS}}^2=(p+Q)^2=(1-x_B)Q^2 \ll Q^2 \ ,
\end{align}
much harder than the intrinsic non-perturbative scale $\Lambda_{\rm QCD}$, but much softer than the hard scale $Q^2$. As a result, in perturbation theory there are large logarithms involving this scale, requiring re-summation. The object that achieves this resummation and separates all the logarithms depending on $\mu_i^2$
is called  the ``jet function''~\cite{Sterman:1986aj,Korchemsky:1992xv,Bosch:2004th,Becher:2006mr,Becher:2006qw} and can be defined as a quark-quark correlator attached to the light-like Wilson-lines along the the direction of incoming collinear quark~\cite{Sterman:1986aj,Korchemsky:1992xv,Becher:2006mr,Becher:2006qw}. Similar to the TMD soft functions~\cite{Collins:1981uk, Ji:2004wu, Echevarria:2015byo,Lubbert:2016rku, Ji:2019ewn,Ji:2020ect}, in the threshold region, soft emissions with $z\ll 1$ can not change the momentum and direction of the fast-moving incoming quark, leading to the light-like Wilson line in the jet function. 

The above picture extends naturally to the threshold limit of matching kernel for the quasi-PDF. Let's consider the 
quasi-PDF for a collinear quark with momenta $p$ in perturbation theory.  When $y \rightarrow 1$, the $z$-component momentum $(1-y)p^z$ flowing from $0$ to $z$ becomes much softer than $p^z$, leading to large threshold logarithms. Furthermore, in the threshold limit, at leading power most of the incoming (out going) collinear momenta simply flow out (in) directly at the quark-link vertices at $0$ and $z$ in Eq.~(\ref{eq:defquasi}), similar to the case of DIS. This is due to the fact that otherwise there will be at least two hard real exchanges with almost opposite $k^z={\cal O}(p^z)$, which is strongly suppressed due to the small phase space available.  As a result, up to the overall phase factor $e^{ip^z z}$,  the {\it intrinsic} $z$ dependency of the quasi-PDF is conjugate to $(1-y)p^z$. When $y \rightarrow 1$, $z$ becomes large, in this way the threshold limit can also be viewed in coordinate space as the {\it large $z$ limit} of the quasi-PDF with the natural ``semi-hard'' scale 
\begin{align}
\mu_i=|1-y|p^z\sim \frac{1}{|z|} \ .
\end{align}
Similar to the case of DIS, all the exchanges at this scale between the incoming/outoing collinear quark and the 
static gauge link in $z$ direction can be re-summed in terms of a simple jet function. Notice that in the current case the anti-collinear quark in DIS has been replaced by a static gauge link in $z$ direction, which can be viewed as a heavy-quark moving in the imaginary time direction. Therefore the jet function is called ``space-like heavy quark jet function'', explaining its name in the title of this section. The space-time picture of the threshold factorization for DIS and quasi-PDF are shown in Fig.~\ref{fig:DISquasi}.

\begin{figure}[h!]
    \centering
    \includegraphics[height=5cm]{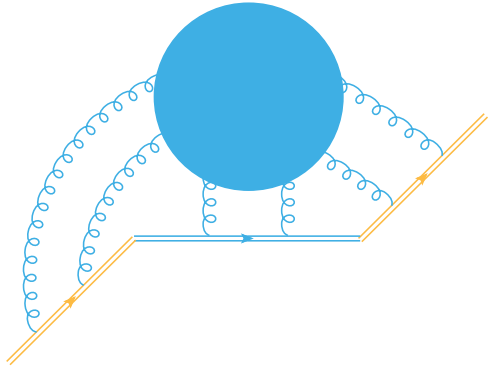}
    \caption{The space-like jet function $\tilde J(\mu z)$. Double lines represent gauge-links. The $t$-direction is vertical while the $z$-direction is horizontal.}
    \label{fig:spacej}
\end{figure}

Inspired by the underlining space-time picture,  we now define the jet function in position space.  Defining the gauge-links
\begin{align}
&W_{n,-}(x)=P\exp \bigg[ig\int_{-\infty}^{0} ds n\cdot A(x+sn)\bigg] \ , \\ 
&W_z(x)=P\exp \bigg[ig\int_{0}^\infty ds n_z\cdot A(x+sn_z)\bigg] \ , \\
&W_{n,+}=P\exp \bigg[ig\int_{0}^{\infty} ds n\cdot A(x+sn)\bigg] \ ,
\end{align}
in terms of which, the space-like jet function can be defined as as the cut-correlator
\begin{align}
\tilde J(\mu z)=\langle \Omega|{\cal \bar T}W_{n,-}^{\dagger}(zn_z)W_z^{\dagger}(zn_z){\cal T}W_z(0)W_{n,-}(0)|\Omega\rangle \ .
\end{align}
Equivalently, it can also be defined without cut as the correlator
\begin{align}\label{eq:deftildeJ}
\tilde J(\mu z)=\langle \Omega|{\cal T}W_{n,+}(zn_z)W_z^{\dagger}(zn_z)W_z(0)W_{n,-}(0)|\Omega\rangle \ .
\end{align}
The second representation has the advantage of being below threshold and time-ordering independent, therefore its high level of analyticity is manifest. Intuitively, it can be viewed as a {\it transition form-factor} where an incoming light-like heavy-quark travelling at velocity of light transits into a space-like heavy-quark, traveling for an imaginary time $z$, then transits back into the outgoing light-like one. The only scale of the jet function is $|z|$, conjugating to $|1-y|p^z$. We will use the form-factor representation for the calculation. See Fig.~\ref{fig:spacej} for a depiction of $\tilde J(\mu z)$.

\subsection{Total result at NLO and RGE in coordinate space}

At leading order, the jet function simply equals to identity. At NLO, the jet function is calculated in Appendix. \ref{sec:NLOjet} in dimensional regularization (DR) with $D=4-2\epsilon$. The results reads
\begin{align}
\tilde J(\mu z,D)=1+\frac{g^2 C_F\Gamma(\frac{D}{2}-1)(\mu_0 |z|)^{4-D}}{4\pi^{\frac{D}{2}}}\bigg(\frac{2}{(D-4)^2}-\frac{1}{(D-4)(D-3)}\bigg) \ .
\end{align}
Now, introducing the standard replacement $\mu^2 =4\mu_0^2 e^{-\gamma_E}\pi$ for $\overline{\text MS}$ scheme, one has the renormalized one-loop jet function in coordinate space after expanding around $\epsilon=0$ as
\begin{align}\label{jetoneloop}
\tilde J(\mu z)=\tilde J(\mu^2 z^2)=1+\frac{\alpha_s C_F}{2\pi} \bigg(\frac{1}{2}l_z^2+l_z+\frac{\pi^2}{12}+2\bigg) \ ,
\end{align}
where $\alpha_s=\frac{g^2}{4\pi}$, and we have defined 
\begin{align}\label{eq:deflz}
l_z=\ln \frac{e^{2\gamma_E}\mu^2z^2}{4} \ .
\end{align}
Notice that the space-like jet function depends only on $|z|$ and is real. This can be argued to all orders in the following way. In the time-ordered perturbation theory, the incoming and outgoing light-like gauge links are represented as heavy quarks travelling at speed of light with dispersion relation $E=k^z$, as a result, all the energy denominators are of the same sign and can not produce any imaginary part. The only source for imaginary part is therefore the gauge link in $z$ direction, represented as a static operator at $t=0$. One can see immediately that flipping the direction of gauge link simply changes the sign of the imaginary part. However, for $z>0$ the Wilson-loop in Eq.~(\ref{eq:deftildeJ}) contains only space-like and light-like separations and is below threshold, therefore real. This implies that for $z<0$ the result is also real. This argument is similar to that given in Appendix A of Ref.~\cite{Ji:2021znw} for imaginary part of quasi-LFWF amplitudes.

Similar to the time-like heavy-quark jet function in Ref.~\cite{Jain:2008gb}, the space-like jet function in coordinate space satisfies a simple renormalization group equation
\begin{align}\label{eq:RGEtiledJ}
\frac{d \ln \tilde J(\mu^2 z^2)}{d\ln \mu}=\Gamma_{\rm cusp}(\alpha_s) l_z-\tilde \gamma_{J}(\alpha_s) \ .
\end{align}
The $\Gamma_{\rm cusp}(\alpha_s)$ is the standard light-like cusp-anomalous dimension~\cite{Korchemsky:1987wg} and is known up to four-loops~\cite{Henn:2019swt,vonManteuffel:2020vjv}, while $\tilde \gamma_{J}(\alpha_s)$ is the same as the anomalous dimension for the ``heavy-quark jet function'' in Ref.~\cite{Jain:2008gb}. In fact, it can be expressed in terms of the UV anomalous dimension $-2\gamma_{HL}$ of a heavy-light Wilson-line cusp originally calculated in Ref.~\cite{Korchemsky:1992xv} and the ``soft-anomalous dimension'' $\gamma_s$ for a light-light Wilson-line cusp~\cite{Korchemskaya:1992je}. 
One can check that 
\begin{align}
\tilde \gamma_{J}=2\gamma_{HL}-2\gamma_s \ .
\end{align}
Notice that the soft anomalous dimension $\gamma_s$ comes from the implicit ``zero-bin subtraction''~\cite{Becher:2006mr, Manohar:2006nz, Jain:2008gb} when calculating the jet function. Given these, it is easy to show that 
\begin{align}\label{eq:resulttildegammaJ}
\tilde \gamma_{J}(\alpha_s)=-\frac{\alpha_s C_F}{\pi}+\left(\frac{\alpha_s}{4\pi}\right)^2\bigg[C_FC_A\left(-\frac{1396}{27}+\frac{23\pi^2}{9}+20\zeta_3\right)+C_Fn_f\left(\frac{232}{27}-\frac{2\pi^2}{9}\right)\bigg] \ .
\end{align}
As expected, the above reproduces the same anomalous dimension for the heavy-quark jet function defined through time-like Wilson lines in Ref.~\cite{Jain:2008gb}. 

\subsection{Beyond NLO: relation to the time-like jet function}

\begin{figure}[h!]
    \centering
    \includegraphics[height=5cm]{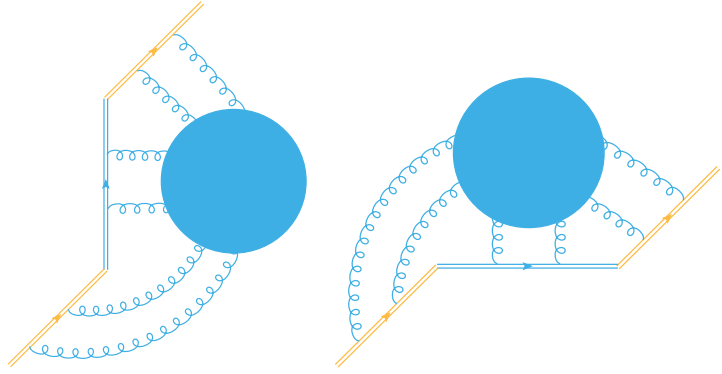}
    \includegraphics[height=4cm]{space-like.eps}
    \caption{The time-like jet function $\tilde J(\mu t)$ (left) and the space-like jet function $\tilde J(\mu z)$ (right) . The $t$-direction is vertical while the $z$-direction is horizontal. The time-like jet function simply relates to the space-like one through changing $t\rightarrow -i|z|$ when $t>0$ up to all orders in PT. }
    \label{fig:timespacej}
\end{figure}

Beyond NLO, in order to utilize the known results in the literature, one needs another less-trivial fact about the space-like jet function: it relates to a time-like version $\tilde J(\mu t)$ with time-like heavy gauge-link simply through $t\rightarrow -i|z|$ when $t>0$:
\begin{align}
\tilde J(\mu t)=\langle \Omega|{\cal T}W_{n,+}(t n_t)W_t^{\dagger}(t n_t)W_t(0)W_{n,-}(0)|\Omega\rangle \ ,
\end{align}
where $W_{t}(x)$ is defined in a way similar to $W_z(x)$ with $n_z$ replaced by $n_t=(1,0,0,0)$. See Fig.~\ref{fig:timespacej} for a depiction of the time-like jet function. This fact holds to all orders in PT and we will provide two proofs for this fact in Appendix \ref{sec:relation}.  On the other hand, at NNLO as claimed in Appendix A of Ref.~\cite{Jain:2008gb}, the light-like gauge link from $-\infty n$ to $0$ in $\tilde J(\mu t)$ can also be chosen from $\infty n$ to $0$, due to the fact that all the resulting differences are scaleless. This observation allows the authors of Ref.~\cite{Jain:2008gb} to calculate their target heavy-quark jet function based on relatively easier Feynman-integrals for $\tilde J(\mu t)$. In fact, we have checked that all the integrals in Eq. (A5) and Eq. (A6) in Ref.~\cite{Jain:2008gb} correspond exactly to our $\tilde J(\mu t)$ and one can indeed flip the direction of the incoming light-like gauge-link. Given this, it is easy to show using spectral representation that our $\tilde J(\mu |z|)$, at NNLO equals to $mB(y,\mu)$ in Ref.~\cite{Jain:2008gb} with $y\rightarrow -i|z|$. More precisely, one needs Eq.~(53) in Ref.~\cite{Jain:2008gb}.

\subsection{Space-like jet function in momentum space}
To check the threshold logarithm, one needs the momentum space version of the jet function. This can be done using the following relations for the Fourier transform ($\lambda=z\cdot p=zp^z$)
\begin{align}\label{eq:defprin}
&\frac{1}{2\pi}\int_{-\infty}^{\infty} d\lambda e^{i\lambda y}\ln \frac{\mu^2e^{2\gamma_E}\lambda^2}{4p_z^2}=-{\cal P}\left(\frac{1}{|y|}\right)-\ln \frac{4p_z^2}{\mu^2}\delta(y) \ , \\ 
&\frac{1}{2\pi}\int_{-\infty}^{\infty} d\lambda e^{i\lambda y}\ln^2 \frac{e^{2\gamma_E}\mu^2\lambda^2}{4p_z^2}=2{\cal P}\bigg(\frac{\ln \frac{4p_z^2y^2}{\mu^2}}{|y|}\bigg)+\bigg(\frac{\pi^2}{3}+\ln^2 \frac{\mu^2}{4p_z^2}\bigg)\delta(y) \ .
\end{align}
The principal values for even distributions are defined as
\begin{align}\label{eq: defprin}
   \langle  {\cal P}(\frac{1}{|x|}),\varphi \rangle = \int_{|x|<1} dx \frac{\varphi(x)-\varphi(0)}{|x|}+ \int_{|x|>1} dx \frac{\varphi(x)}{|x|} \ , \\ 
   \langle  {\cal P}(\frac{\ln x^2}{|x|}),\varphi \rangle = \int_{|x|<1} dx \frac{\ln x^2}{|x|}\bigg(\varphi(x)-\varphi(0)\bigg) +\int_{|x|>1} dx \frac{\ln x^2 }{|x|} \varphi(x) \ .
   \end{align}
For old distribution, it agrees with the standard Cauchy principal value. In Appendix \ref{sec:fourier} we will provide a self-contained derivation of these Fourier transformation rules. As a result, one has
\begin{align}\label{eq: jetbare}
&p^z\tilde J\left(\frac{(1-y)p_z}{\mu}\right)=\frac{1}{2\pi}\int_{-\infty}^{\infty} d\lambda e^{-i\lambda (1-y)} \tilde J(\frac{\mu^2}{p_z^2}\lambda^2)\nonumber \\ 
&=\delta(1-y)+\frac{\alpha_sC_F}{2\pi}\bigg[{\cal P}\bigg(\frac{\ln (1-y)^2}{|1-y|}\bigg)+(L_z-1){\cal P}\left(\frac{1}{|1-y|}\right)+\delta(y-1)\bigg(2+\frac{\pi^2}{4}+\frac{1}{2}L_z^2-L_z\bigg)\bigg] \ ,
\end{align}
where one has defined
\begin{align}\label{eq:defLz}
L_z \equiv \ln \frac{4p_z^2}{\mu^2} \equiv \ln \frac{\zeta_z}{\mu^2} \ .
\end{align}
Unfortunately, the $L_z-1$ term differs from that of the one-loop quasi-PDF by a factor of ${\cal P}(\frac{1}{1-y})$, which will be provided by the imaginary part of the hard kernel.

\section{Universal heavy-light Sudakov form factor and redefinition of the jet function}\label{sec:suda}

In the previous section, we have introduced the most crucial object in the threshold re-summation: the space-like jet function resuming all the semi-hard exchanges at scale $\mu_i=|1-y|p^z\sim \frac{1}{z}$ in the large $z$ or large $y$ limit, fluctuating between the collinear quark and the static gauge link.  On the other hand, the incoming collinear quark, when hit by the quark-link vertex, transits into a space-like heavy quark with low virtuality $|1-y|p^z \sim |z|^{-1}$.  This leads to strong Sudakov effect due to virtual fluctuations nearby, similar to the case where a collinear quark becomes another anti-collinear quark in the case of DIS.

As usual, the Sudakov effect can be resummed to all orders using a universal object, the heavy-light Sudakov form factor, which is similar to the standard Sudakov form factor~\cite{Sudakov:1954sw,Collins:1989bt,Moch:2005id,Baikov:2009bg} involving only light-quarks. For quasi-PDF since there are two quark-link vertices, one needs a pair of Sudakov form factors, each one with a light external on-shell quark with $p^2=0$ and a external heavy gauge-link extending to infinity.  The natural scale of the form factor is $\zeta_z=\frac{4|p\cdot n_z|^2}{n_z^2}=4 p_z^2$, 
which is the only scalable quantity that can be formed in terms of Lorentz invariant combinations of $p$ and $n_z$. We have already encountered such object in factorization of quasi-TMDPDF~\cite{Ebert:2019okf,Ji:2019ewn,Ji:2020ect,Vladimirov:2020ofp,Ebert:2020gxr,Ebert:2022fmh} and quasi-LFWF~\cite{Ji:2021znw,Ji:2020ect}. In the current case the hard kernel simply agrees with the quasi-LFWF one
\begin{align}
H_{\rm HL}\left(\frac{\zeta_z}{\mu^2},{\rm sign}(z)\right) \equiv H_{\rm LFWF}^{-{\rm sign}(z)}\left(\frac{\zeta_z}{\mu^2}\right) \ .
\end{align}
Notice that there are imaginary part of each Sudakov form factor, as demonstrated in Ref.~\cite{Ji:2021znw}. In case of quasi-TMDPDF they cancel with each other, while for the threshold limit and for the quasi-LFWF the imaginary parts of the two Sudakov kernels add together and depend on the sign of $z$~\cite{Ji:2021znw}. At one-loop, one has~\cite{Ji:2021znw} 
\begin{align}
&H^{\pm}_{\rm LFWF}=1+\frac{\alpha_s C_F}{4\pi} \bigg(-4-\frac{5\pi^2}{6}+2L_{\pm}-L_{\pm}^2\bigg) \ , \\ 
&L_{\pm}\equiv \ln \frac{-\zeta_z+i0}{\mu^2}=L_z\pm i\pi \ .
\end{align}
More generally, the RGE for the above heavy-light Sudakov form factor reads~\cite{Ji:2019ewn,Ji:2021znw}
\begin{align}\label{eq:RGEH}
\frac{d}{d\ln \mu} \ln H_{\rm HL}\left(\frac{\zeta_z}{\mu^2},\alpha(\mu),{\rm sign}(z)\right)=\Gamma_{\rm cusp}(\alpha)\ln\frac{\zeta_z}{\mu^2}+\tilde {\gamma}_{H}(\alpha)-i\pi{\rm sign}(z)\Gamma_{\rm cusp}(\alpha) \ .
\end{align}
Here $\tilde \gamma_{H} \equiv \gamma_C$ is the same anomalous dimension for the hard kernel of the quasi-TMDPDF factorization and is known to two-loop~\cite{Ji:2019ewn,Ji:2021znw}. We will discuss more on anomalous dimensions in Appendix \ref{sec:anomalousd}. 

Given all above, it is much more convenient to redefine the hard kernel to be purely real, while assign the phase to the jet function:
\begin{align}\label{eq:hardabs}
 &H(L_z) \equiv |H|_{\rm HL}\left(L_z \right)\ , \\
 & J_f(z\mu)\equiv\tilde J(z^2\mu^2) \exp \bigg[i {\rm sign}(z) A(L_z) \bigg] \ , \label{eq:defJf}
\end{align}
where we define the phase angle as
\begin{align}\label{eq:phaseangle}
{\rm sign}(z) A(L_z)={\rm Arg}\bigg(H_{\rm HL}\left(\frac{\zeta_z}{\mu^2},{\rm sign}(z)\right)\bigg).
\end{align}
Notice that the phase angle is proportional to ${\rm sign}(z)$ to all orders. 
After this redefinition, one can see that the real part of $J_f$ at one-loop remains the same as $\tilde J$, while the imaginary part becomes
\begin{align}\label{eq:JetoneloopIM}
{\rm Im} J_f(z\mu)=\frac{\alpha_sC_F}{2\pi}(L_z-1)i\pi {\rm sign}(z) \ .
\end{align}
After Fourier transform, one simply has
\begin{align}
{\cal F}\left(i\pi{\rm sign}(\lambda)\right)(1-y)={\cal P} \bigg(\frac{1}{1-y}\bigg) \ .
\end{align}
Adding this to Eq.~(\ref{eq: jetbare}), one has the final NLO jet function 
\begin{align} \label{eq:jetonloop}
&p^zJ_{f}\left(\frac{(1-y)p^z}{\mu},\frac{4p_z^2}{\mu^2}\right)=\delta(1-y)\bigg[1+\frac{\alpha_sC_F}{2\pi}\left(2+\frac{\pi^2}{4}+\frac{L_z^2}{2}-L_z\right)\bigg]\nonumber \\ 
&+\frac{\alpha_sC_F}{2\pi}\bigg[{\cal P}\left(\frac{\ln (1-y)^2}{|1-y|}\right)+(L_z-1)\bigg({\cal P}\left(\frac{1}{1-y} \right) +{\cal P}\left(\frac{1}{|1-y|} \right)\bigg)\bigg] \ ,
\end{align}
and the NLO hard kernel
\begin{align}\label{hardoneloop}
H(L_z)=1+\frac{\alpha_s C_F}{2\pi}\bigg(-2+\frac{\pi^2}{12}-\frac{L_z^2}{2}+L_z\bigg) \ .
\end{align}
It is easy to check that the above hard kernel together with the jet function $J_f$ completely reproduces the one-loop quasi-PDF in $\overline{\text MS}$ scheme~\cite{Xiong:2013bka,Izubuchi:2018srq} at power $(1-y)^{-1}$ through the factorization formula Eq.~(\ref{eq:threshPT}), when the plus-functions are converted to the principle value using the rules in Appendix \ref{sec:fourier}. Moreover, the hard kernel completely agrees with the hard kernel for quasi-TMDPDF/quasi-LFWF factorization~\cite{Ebert:2019okf,Ji:2019ewn,Vladimirov:2020ofp,Ebert:2020gxr,Ji:2021znw}.

\section{Threshold limit at NNLO: explicit results and extraction of the heavy-light Sudakov form factor}\label{sec:NNLO}
In this section, armed with the factorization formalism Eq.~(\ref{eq:threshPT}) and all the ingredients defined in previous sections, we investigate the threshold factorization at NNLO to demonstrate the correctness of our formalism.  Since our jet function agrees with the analytic continuation from the time-like version in Ref.~\cite{Jain:2008gb}, we can completely predict the threshold limit of the $\overline {\text MS}$ matching kernel at power $(1-y)^{-1}$ up to two unknown constants $c_a$ and $c_H$. We show that our prediction completely agrees with the threshold limit extracted directly form the full two-loop matching kernel~\cite{Li:2020xml,Chen:2020ody}. Furthermore, it allows us to extract the two unknown constants $c_a$ and $c_H$ and completely determines the heavy-light Sudakov hard kernel at NNLO that also appears in the quasi-TMDPDF/TMDWF factorization. 

\subsection{NNLO threshold limit in coordinate space}\label{sec:thresholcord}
Here we present the results of the quark quasi-PDF in the threshold limit, predicted by our factorization formula Eq.~(\ref{eq:threshPT}). Our notation are $H=1+\alpha_s H^{(1)}+\alpha_s^2 H^{(2)}$, $\tilde J = 1 + \alpha_s \tilde J^{(1)} + \alpha_s^2 \tilde J^{(2)}$, $J_f=1+\alpha_s J_f^{(1)}+\alpha_s^2 J_f^{(2)}$, $\gamma=\gamma^{(1)}\alpha_s+\gamma^{(2)}\alpha_s^2$ and ${\rm Arg}\bigg(H_{\rm HL}\left(\frac{\zeta_z}{\mu^2},{\rm sign}(z)\right)\bigg)={\rm sign}(z) A ={\rm sign}(z) A^{(1)}(L_z) \alpha_s + {\rm sign}(z) A^{(2)}(L_z) \alpha_s^2$. We first consider the space-like jet function. Using the RGE Eq.~(\ref{eq:RGEtiledJ}) of the jet function, one has
\begin{align}\label{jettwoloop}
\ln \tilde J^{(2)}=\frac{C_F\beta_0}{24\pi}l_z^3+\frac{1}{4}\left(\Gamma^{(2)}_{\rm cusp}+\frac{C_F\beta_0}{2\pi}\right)l_z^2+\bigg(\frac{C_F\beta_0}{4\pi}\left(2+\frac{\pi^2}{12}\right)-\frac{\tilde \gamma_{J}^{(2)}}{2}\bigg)l_z+c_1 \ .
\end{align}
Here $\Gamma_{\rm cusp}^{(2)}$ is given in Eq.~(\ref{eq:resultcusp}), and $\tilde \gamma_J^{(2)}$ is given in Eq.~(\ref{eq:resulttildegammaJ}). As we show in Appendix~\ref{sec:relation}, the constant term $c_1$ is the same as that in the time-like jet function~\cite{Jain:2008gb}
\begin{align}\label{eq:resultc1}
c_1=\frac{2\beta_0C_F}{\pi}\bigg(\frac{281}{216}+\frac{5\pi^2}{576}-\frac{\zeta_3}{48}\bigg)+\frac{C_FC_A}{\pi^2}\bigg(-\frac{11}{54}+\frac{7\pi^2}{144}-\frac{17\pi^4}{2880}-\frac{5\zeta_3}{8}\bigg) \ .
\end{align} 
Notice $l_z$ is defined in Eq.~(\ref{eq:deflz}), and our convention for $\beta_0$ is
\begin{align}\label{eq:beta}
    \frac{d\alpha}{d\ln \mu}=\beta =-\beta_0 \alpha^2 - \beta_1 \alpha^3 +... \ ,  \ 
    \beta_0=\frac{11C_A}{6\pi}-\frac{n_f}{3\pi} \ .
\end{align}
One also needs the two-loop value of the imaginary part\label{eq:jet ima}
\begin{align}
A^{(2)}(L_z)=\frac{\pi}{2}\bigg[\Gamma^{(2)}_{\rm cusp}L_z-\frac{C_F\beta_0}{2\pi}\left(\frac{L_z^2}{2}-L_z\right)+c_a\bigg] \ ,
\end{align}
where $L_z$ is defined in Eq.~(\ref{eq:defLz}). Combined together, one has
\begin{align}\label{eq:resultJf2}
J_{f}^{(2)}=\frac{1}{2}(J_f^{(1)})^2+\ln \tilde J^{(2)}+i {\rm sign}(z) A^{(2)}(L_z) \ , \\ 
J_f^{(1)}=\frac{C_F}{2\pi}\left(\frac{l_z^2}{2}+l_z+\frac{\pi^2}{12}+2\right)+i{\rm sign}(z) A^{(1)}(L_z) \ , \label{eq:resultJf1}
\end{align}
where $A^{(1)}(L_z)= \frac{C_{F}}{2}(L_z-1)$ is given in Eq.~(\ref{eq:JetoneloopIM}).
Finally, one also has the two-loop hard kernel
\begin{align}\label{hardtwoloop}
\ln H^{(2)}=\frac{C_F\beta_0}{24\pi}L_z^3-\frac{1}{4}\left(\Gamma^{(2)}_{\rm cusp}+\frac{C_F\beta_0}{2\pi}\right)L_z^2-\bigg(\frac{C_F\beta_0}{4\pi}\left(-2+\frac{\pi^2}{12}\right)+\frac{\tilde \gamma_{H}^{(2)}}{2}\bigg)L_z+c_H \ .
\end{align}
The explicit expression for $\tilde \gamma_H^{(2)}$ is given in Eq.~(\ref{eq:tildegammaHresult}). 
Given all the above, our two-loop prediction for the quasi-PDF in the threshold limit reads in coordinate space as
\begin{align}
\tilde f^{(2)}(z,L_z)\rightarrow e^{-izP^z} f^{(2)}(l_z,{\rm sign}(z), L_z) \ , 
\end{align}
where one has
\begin{align}
 f^{(2)}=J_f^{(2)}+\ln H^{(2)}+H^{(1)}J_f^{(1)}+\frac{1}{2}(H^{(1)})^2 \ .
\end{align}
When Fourier-transforming (the same convention as Eq.~(\ref{eq: jetbare})) it into momentum space, one obtains all the singular terms $\delta(1-y)$, $\frac{1}{y-1}$, $\frac{\ln (y-1)}{y-1}$, $\frac{\ln^2(y-1)}{y-1}$, $\frac{\ln^3(y-1)}{y-1}$. There are 41 non-vanishing terms, all of them are consistent with the exact results. Detailed expressions are presented in Appendix \ref{sec: NNLOmomemtum}. In particular, this confirms the correctness of our $\tilde \gamma_J$ and $\tilde \gamma_H$. 

\subsection{Extraction of $c_a$ and $c_H$. Complete determination of the universal heavy-light Sudakov form factor $H$}
Through comparing our prediction for the $\frac{1}{1-y}$ term based on threshold factorization in either $y>1$ or $y<1$ region with the exact calculation~\cite{Li:2020xml,Chen:2020ody}, one confirms the correctness of $\tilde \gamma_J$, and extracts $c_a$ as
\begin{align}\label{eq:resultca}
c_a = \left(-\frac{3 \zeta_3}{\pi ^2}+\frac{7}{12}-\frac{1}{2 \pi ^2}\right) C_F^2 + \left(\frac{11 \zeta_3}{4 \pi ^2}-\frac{11}{24}-\frac{475}{108 \pi ^2}\right) C_F C_A + \left(\frac{1}{6}+\frac{38}{27 \pi ^2}\right) C_F n_f T_F
\end{align}
Through comparing the $\delta(1-y)$ term based on threshold factorization with the exact calculation~\cite{Li:2020xml,Chen:2020ody}, one confirms the correctness of all the anomalous dimensions, and obtains the constant term in the two loop hard kernel $\ln H^{(2)}$ as 
\begin{align}\label{eq:resultcH}
c_H= \ \ &\bigg(\frac{241 \zeta_3}{144 \pi ^2}+\frac{11 \pi ^2}{320}-\frac{559}{1728}-\frac{971}{324 \pi ^2}\bigg)C_FC_A+\bigg(\frac{-45 \zeta_3-2 \pi ^4+30 \pi ^2-30}{24 \pi ^2}\bigg)C_F^2 \nonumber \\ 
+&\bigg(\frac{36 \zeta_3+51 \pi ^2+1312}{1296 \pi ^2}\bigg)C_Fn_fT_F \ .
\end{align}
Numerically, one has 
\begin{align}
c_H = 0.0725 \, C_F^2 - 0.0840 \, C_F C_A + 0.1453 \, C_F n_f T_F \ .
\end{align}
Details of the extraction of $c_H$ are presented in Appendix \ref{sec:convert} and \ref{sec:cHanaly}.  Combining with $c_a$, the above completely determines the heavy-light Sudakov form factor at two-loop, a universal object that also appears in TMD factorization for quasi-TMDPDFs and quasi-LFWFs.

\section{Evolution equations and RGE resummation for the matching kernel}\label{sec: evo}
To summarize, the perturbative quark quasi-PDF for an incoming collinear quark with momentum $p$, in the threshold limit reads
\begin{align}
\tilde f\left(y,\frac{\zeta_z}{\mu}\right)=H\left(\frac{4p_z^2}{\mu^2}\right)p^zJ_f\left(\frac{(1-y)p^z}{\mu},\frac{4p_z^2}{\mu^2}\right)\left(1+{\cal O}(1-y)\right) \ ,
\end{align}
to all orders in perturbation theory. The matching kernel~\cite{Ji:2014gla,Xiong:2013bka,Izubuchi:2018srq}, in the $\xi=\frac{y}{x}\rightarrow 1$ limit can then be read directly by the substitution $y\rightarrow \xi$ and $p^z \rightarrow |xP^z|$, leading to Eq.~(\ref{eq:finalmatchingC}).
Clearly, after using the RGE, the scale in the jet function, $\mu_i=|1-\xi||xP^z|$ will be re-summed, leading to all the threshold logarithms. We now collect all the RGE and show that after using the relations of the anomalous dimensions, the matching kernel matches precisely with the threshold limit of the DGLAP kernel. This allows us to use the formalism in Ref.~\cite{Becher:2006mr} to obtain a fully RGE re-summed form of the matching kernel in momentum space, facilitating further applications to lattice calculation. 

\subsection{Evolution equation and matching to DGLAP}
As a reminder, for the absolute value of the hard kernel, the renormalization group equation reads
\begin{align}
\frac{d}{d\ln \mu} \ln H\left(\frac{\zeta_z}{\mu^2},\alpha(\mu)\right)=\Gamma_{\rm cusp}(\alpha)\ln\frac{\zeta_z}{\mu^2}+\tilde {\gamma}_{H}(\alpha) \ .
\end{align}
On the other hand, the evolution equation for the jet function in momentum space reads
\begin{align}
&\frac{d}{d\ln \mu} J_{f}\left(\frac{(1-\xi)xP^z}{\mu}\right)=-\left(\tilde \gamma_{J}(\alpha_s)+\Gamma_{\rm cusp}(\alpha_s)\ln \frac{4x^2 P_z^2}{\mu^2}\right)J_f\left(\frac{(1-\xi)xP^z}{\mu},\alpha(\mu)\right)
\nonumber \\ 
&-2\Gamma_{\rm cusp}(\alpha_s){\cal P}\int_{\xi'>\xi} \frac{J_{f}\left(\frac{(1-\xi')x P^z}{\mu},\alpha(\mu)\right)}{\xi'-\xi}d\xi' \ .\end{align}
Combining the above, one has
\begin{align}\label{eq:evoC}
\frac{d}{d\ln \mu}{\cal C}\left(\xi,\frac{xP^z}{\mu}\right)=(\tilde \gamma_{H}-\tilde \gamma_{J}){\cal C}\left(\xi,\frac{xP^z}{\mu}\right)-2\Gamma_{\rm cusp}{\cal P}\int_{\xi'>\xi} \frac{{\cal C}\left(\xi', \frac{xP^z}{\mu}\right)}{\xi'-\xi}d\xi' \ ,
\end{align}
which we now show to agree with the DGLAP evolution equation in the endpoint region with splitting fraction $z\rightarrow 1$.

For this purpose, one needs the universal endpoint limit of the splitting function~\cite{Moch:2004pa,Becher:2006mr}
\begin{align}
P_{qq}(z)|_{z\rightarrow 1}=\frac{2\Gamma_{\rm cusp}}{(1-z)^+}+2\gamma_{\phi} \delta(1-z) \ .
\end{align}
It is easy to see that the second term in Eq.~(\ref{eq:evoC}) simplify matches with the $\frac{2\Gamma_{\rm cusp}}{(1-z)^+}$. In order for the $\delta(1-y)$ term to match, one must have
\begin{align}\label{eq:anrela}
2\gamma_{\phi}+\tilde \gamma_{H}-\tilde \gamma_{J}\equiv 2\gamma_{\phi}+\tilde \gamma_{H}-2\gamma_{HL}+2\gamma_s=2\gamma_F \ ,
\end{align}
where $\gamma_F$ is the UV anomalous dimension of a heavy-light current~\cite{Ji:1991pr, Chetyrkin:2003vi}. However, by studying the threshold factorization of DIS in terms of quark jet function, one can show that $\gamma_{\phi}$ is simply related to the soft anomalous dimension $\gamma_s$ and the anomalous dimension $\gamma_V$ of the light-light Sudakov hard kernel~\cite{Moch:2005id} through
\begin{align}
\gamma_{V}+\bigg(\gamma_{\phi}+\gamma_s\bigg)+\gamma_{\phi}=0 \ ,
\end{align}
where $\gamma_{\phi}+\gamma_{s}$ is just $-\gamma_{J}$~\cite{Becher:2006mr,Becher:2006qw}, the constant term of the anomalous dimension for light-quark jet function. As a result, one has $2\gamma_{\phi}=-\gamma_s-\gamma_V$. Eliminating $\gamma_{\phi}$ from Eq.~(\ref{eq:anrela}), one ends up at the relation for $\tilde \gamma_{H}$
\begin{align}
\tilde \gamma_{H}=\gamma_V+2\gamma_{\rm F}+\bigg(2\gamma_{HL}-\gamma_s\bigg) \ .
\end{align}
To show this agrees with our previous formula $\gamma_C$ for quasi-TMDPDF factorization, one simply notice that $\gamma_V=2\gamma_H$, where $\gamma_H$ is the ``hard anomalous dimension'' related to the TMDPDFs through~\cite{Ebert:2019okf,Luo:2019hmp}
\begin{align}
\frac{d}{d\ln \mu}f^{\rm TMD}(x,b_\perp,\mu,\zeta)=\Gamma_{\rm cusp}\ln \frac{\mu^2}{\zeta}-2\gamma_H \ .
\end{align}
which leads to
\begin{align}
\tilde \gamma_H=2\gamma_H+2\gamma_F+(2\gamma_{HL}-\gamma_s) \ .
\end{align}
In fact, the $2\gamma_{HL}-\gamma_s$ in the bracket is nothing but the anomalous dimension $\Gamma_S$ for the ``reduced soft factor''~\cite{Ji:2019ewn,Li:2020xml,Ji:2021znw} or equivalently the ``instant-jet TMD distribution''~\cite{Vladimirov:2020ofp} defined through a ratio between two light-heavy Wilson-loops and one light-light Wilson loop~\cite{Ji:2021znw}. It can also be defined purely in terms of a heavy-heavy Wilson loop at large rapidity gap, using the relation for the large-$Y$ asymptotics for the hyperbolic-angle dependent cusp anomalous dimension~\cite{Korchemsky:1987wg,Grozin:2015kna}
\begin{align}
&\Gamma_{\rm cusp}(\alpha_s,Y)\rightarrow Y\Gamma_{\rm cusp}(\alpha_s)+\gamma_{HH}(\alpha_s) \ , \\ 
&\gamma_{HH}\equiv \Gamma_S=2\gamma_{HL}-\gamma_s \ .
\end{align}
The above relation will be verified explicitly in Appendix~\ref{sec:anomalousd}.

\subsection{RGE resummation of the matching kernel}
In this subsection we will use the renormalization group equations for the hard kernel and the jet function to perform the resummation of the matching kernel ${\cal C}$. Our resummation strategy is similar to that in Ref.~\cite{Becher:2006mr}.

We first consider RGE resummation of the hard kernel $H$. Using the renormalization group equation Eq.~(\ref{eq:RGEH}) and the evolution factors defined in Eq.~(\ref{eq:aGamma}) and Eq.~(\ref{eq:othera}), one evolves from the hard scale $\zeta_z$ to $\mu$ and the resumed form of the hard kernel reads
\begin{align}
&H\left(\frac{\zeta_z}{\mu^2},\alpha(\mu)\right)=H(\alpha(\zeta_z))\exp \bigg[2S(\zeta_z,\mu)-a_{H}(\zeta_z,\mu)\bigg] \ , \\ 
&A\left(\frac{\zeta_z}{\mu^2}\right)=A\left(\alpha(\zeta_z)\right)+\pi a_{\Gamma}(\zeta_z,\mu) \ , \label{eq:phaseRG}
\end{align}
where $H(\alpha(\zeta_z))$ is the absolute value of hard kernel defined in Eq.~(\ref{eq:hardabs}) and $A\left(\alpha(\zeta_z)\right)$ is the phase angle defined in Eq.~(\ref{eq:phaseangle}).

We then consider the RGE resummation of the space-like jet function. We start from the coordinate space jet function defined in Eq.~(\ref{eq:deftildeJ}). Based on the RG equation Eq.~(\ref{eq:RGEtiledJ}), and the evolution factors defined in Eq.~(\ref{eq:aGamma}) and Eq.~(\ref{eq:othera}), one evolves from the semi-hard scale $\mu_i$ to scale $\mu$,
\begin{align}
\tilde J\left(\mu^2z^2,\alpha(\mu)\right)=\exp\bigg[-2S(\mu_i,\mu)+a_{J}(\mu_i,\mu)\bigg]\bigg(\frac{4}{z^2e^{2\gamma_E}\mu_i^2}\bigg)^{a_{\Gamma}(\mu_i,\mu)}\tilde J(l_z(\mu_i),\alpha(\mu_i)) \ ,
\end{align}
for the jet function without phase. Absorbing all the phases in Eq.~(\ref{eq:phaseRG}), one has
\begin{align}\label{eq:JetRGCoor}
J_{f}(\mu z,\alpha(\mu),\zeta_z)=&\exp\bigg[i{\rm sign}(z)\left(\pi a_{\Gamma}(\zeta_z,\mu)+A(\alpha(\zeta_z))\right)
\bigg]\nonumber \\ 
&\times \exp\bigg[-2S(\mu_i,\mu)+a_{J}(\mu_i,\mu)\bigg]\bigg(\frac{2}{|z|\mu_ie^{\gamma_E}}\bigg)^{2a_{\Gamma}(\mu_i,\mu)}\tilde J(l_z(\mu_i),\alpha(\mu_i)) \ .
\end{align}
Clearly, the scale $\zeta_z=4x^2P_z^2$ coming from the imaginary part is separated from the threshold scale $\mu_i$ that will be chosen as $\mu_i=|(x-y)P^z|=|(1-\xi)xP^z|$. 
We then Fourier transform it to momentum space ($\eta=2a_{\Gamma}(\mu_i,\mu)$) based on Eqs.~(\ref{eq:FTeven}) and~(\ref{eq:FTodd})
\begin{align}
&J_f\left(\frac{p}{\mu},\frac{\zeta_z}{\mu^2},\alpha(\mu)\right)=\exp\bigg[-2S(\mu_i,\mu)+a_{J}(\mu_i,\mu)\bigg]  \nonumber \\ &
\times \tilde J\left(l_z=-2\partial_{\eta},\alpha(\mu_i)\right)\left[\frac{\sin \bigg(\hat A(\zeta_z,\mu){\rm sign}(p)+\frac{\eta \pi}{2}\bigg)}{|p|}\left(\frac{2|p|}{\mu_i}\right)^\eta\right]_{*}\frac{ \Gamma(1-\eta) \mathrm{e}^{-\eta \gamma_E}}{\pi} \ ,
\end{align}
where we write $\hat A(\zeta_z,\mu)=\pi a_{\Gamma}(\zeta_z,\mu)+A(\alpha(\zeta_z))$ and $p=(1-\xi) p_z$ for abbreviation. 
$\tilde J\left(l_z=-2\partial_{\eta},\alpha(\mu_i)\right)$ denotes the fixed order jet function in coordinate space and its argument $\partial_\eta$ denotes a derivative with respect to $\eta$ to generate the log terms, which is the same trick in~\cite{Becher:2006mr}. The star distribution is defined as
\begin{align}\label{eq:star}
\int_{-\infty}^{+\infty} dp \left[\frac{1}{|p|}\left(\frac{2|p|}{\mu_i}\right)^\eta\right]_{*}f(p) = \int_{-\infty}^{+\infty} dp \left[\frac{1}{|p|}\left(\frac{2|p|}{\mu_i}\right)^\eta\right]\left(f(p)-\sum_{k=0}^{n}\frac{f^{(k)}(0)}{k!}p^k\right),
\end{align}
where $\eta\leq0$ and $n$ is an integer for $n\leq-\eta<n+1$. For $\eta>0$, the subtraction terms involving $f^{(k)}(0)$ are not required.

Combining all above, the threshold resummed form of the matching kernel in momentum space reads as
\begin{align}
&{\cal C}\left(\xi,\frac{xP^z}{\mu}\right)=H\left(\frac{4x^2P_z^2}{\mu^2},\alpha(\mu)\right) |xP^z| J_f\left(\frac{(1-\xi)xP^z}{\mu},\frac{4x^2P_z^2}{\mu^2},\alpha(\mu)\right) \nonumber \\ 
&=H(\alpha(\zeta_z))\exp \bigg[2S(\zeta_z,\mu)-a_{H}(\zeta_z,\mu)-2S(\mu_i,\mu)+a_{J}(\mu_i,\mu)\bigg] \nonumber \\
&\times \tilde J\left(l_z=-2 \partial_\eta,\alpha(\mu_i)\right) \left[\frac{\sin \left(\hat A(\zeta_z,\mu){\rm sign}(1-\xi)+\frac{\eta \pi}{2}\right)}{|1-\xi|}\left(\frac{2|1-\xi||xP^z|}{\mu_i}\right)^\eta\right]_{*}\frac{ \Gamma(1-\eta) \mathrm{e}^{-\eta \gamma_E}}{\pi} \ ,
\end{align}
where $\zeta_z=4x^2P_z^2$ and $\mu_i=|(x-y)P^z|=|(1-\xi)xP^z|$. The formula is valid up to ${\cal O}(1-\xi)$ corrections.

\section{Summary and outlook}\label{sec:conc}
In this paper we have presented all the necessary results on the threshold limit of quark quasi-PDF. We have shown that in the threshold limit, the quasi-PDF factorizes through a space-like jet function and a pair of heavy-light Sudakov form factors. By comparing the prediction based on our factorization formula with the exact calculation in Ref.~\cite{Li:2020xml,Chen:2020ody}, we are able to confirm the correctness of our formalism explicitly up to NNLO, and extract the universal heavy-light Sudakov form factor at NNLO.
Using the RGE of the individual pieces, we obtain the resummed form of the matching kernel in momentum space that will be applied to numerical calculations. Finally, the results reveal a high level of universality across the threshold limit of quasi-PDF and the large momentum expansion of quasi-TMDPDFs and LFWFs.

In a future work, we will study further the RGE re-summation of the threshold region of matching kernel based on our current formalism and apply it to numerical determination of pion PDF. 

Note: During the writing-up stage of this paper, we noticed that Ref.~\cite{delRio:2023pse} explicitly computed two-loop heavy-light Sudakov form factor 
in the context of quasi-TMDPDF large-momentum expansion, which 
agrees with the result we found here. 

\acknowledgments
We thank Iain Stewart for pointing out possible connection between time-like and space-like heavy-quark jet functions, and the former
has been computed in Ref~\cite{Jain:2008gb}. All calculations in the paper are cross checked between Y.L and Y.S. This research is supported by the U.S. Department of Energy, Office of Science, Office of Nuclear Physics, under contract number DE-SC0020682.
Y. L. is supported by the Priority Research Area SciMat and DigiWorlds under the program Excellence Initiative - Research University at the Jagiellonian University in Krak\'{o}w. Y.S. is partially supported by the U.S.~Department of Energy, Office of Science, Office of Nuclear Physics, contract no.~DE-AC02-06CH11357.

\appendix
\section{Universality of anomalous dimensions}\label{sec:anomalousd}
 In fact, in all the TMD factorization and threshold factorization formalism for light-light and light-heavy current collerators, besides the universal light-like cusp anomalous dimension $\Gamma_{\rm cusp}(\alpha_s)$~\cite{Korchemsky:1987wg} and the heavy-light current UV anomalous dimension $\gamma_F$~\cite{Ji:1991pr, Chetyrkin:2003vi}, there exists only three independent UV anomalous dimensions serving as basic building blocks, generating all others through linear combinations. We chose them as $\gamma_V$ (for light-light Sudakov hard kernel~\cite{Moch:2005id,Becher:2006mr}), $\gamma_s$ (for light-light Wilson line cusp~\cite{Korchemskaya:1992je}) and $\gamma_{HL}$ (for heavy-light Wilson line cusp~\cite{Korchemsky:1992xv}). Up to two loops, 
the light-like cusp anomalous dimension is given by
\begin{align}\label{eq:resultcusp}
\Gamma_{\rm cusp}=\frac{\alpha_sC_F}{\pi}+\frac{\alpha_s^2}{\pi^2}\bigg[\left(\frac{67}{36}-\frac{\pi^2}{12}\right)C_A C_F-\frac{5}{18}C_Fn_f\bigg] \ ,
\end{align}
while the heavy-light current anomalous dimension $\gamma_F$ is given as~\cite{Ji:1991pr, Chetyrkin:2003vi}
\begin{align}
 \gamma_F=\frac{3\alpha_sC_F}{4 \pi }+ \left(\frac{\alpha_s}{4 \pi }\right)^2 \bigg[\left(\frac{49}{6}-\frac{2 \pi ^2}{3}\right)C_AC_F-\left(\frac{5}{2}-\frac{8 \pi ^2}{3}\right)C_F^2-\frac{5}{3}C_Fn_f\bigg]\ .
 \end{align}
The $\gamma_V$ for light-quark Sudakov form factor reads~\cite{Moch:2005id,Becher:2006mr}
 \begin{align}
 &\gamma_V=-\frac{3C_F\alpha_s}{2\pi} \nonumber \\ +&\left(\frac{\alpha_s}{4\pi}\right)^2\bigg[C_FC_A\left(-\frac{961}{27}-\frac{11\pi^2}{3}+52\zeta_3\right)+C_F^2\left(-3+4\pi^2-48\zeta_3\right)+C_Fn_f\left(\frac{130}{27}+\frac{2\pi^2}{3}\right)\bigg] \ ,
 \end{align}
 notice that the presence of $C_F^2$ term, indicating that this is not for a Wilson loop. For soft anomalous dimension $\gamma_s$ one has~\cite{Korchemskaya:1992je}
 \begin{align}
\gamma_s=\left(\frac{\alpha_s}{4\pi}\right)^2\bigg[C_FC_A\left(\frac{808}{27}-\frac{11\pi^2}{9}-28\zeta_3\right)+C_Fn_f\left(-\frac{112}{27}+\frac{2\pi^2}{9}\right)\bigg] \ ,
\end{align}
notice the absence of $\alpha_s$ order contribution, as well as absence of $C_F^2$ term. For the heavy-light Wilson-line cusp anomalous dimension $\gamma_{HL}$ one has~\cite{Korchemsky:1992xv}
\begin{align}
\gamma_{HL}=-\frac{C_F\alpha_s}{2\pi}+\left(\frac{\alpha_s}{4\pi}\right)^2\bigg[C_FC_A\left(\frac{110}{27}+\frac{\pi^2}{18}-18\zeta_3\right)+C_Fn_f\left(\frac{4}{27}+\frac{\pi^2}{9}\right)\bigg] \ .
\end{align}
Again, $C_F^2$ term is absent. In terms of the above, one can directly show that
\begin{align}
2\gamma_{HL}-\gamma_s=-\frac{\alpha_sC_F}{\pi}+\frac{\alpha_s^2}{\pi^2}\bigg[C_FC_A\left(-\frac{49}{36}+\frac{\pi^2}{12}-\frac{1}{2}\zeta_3\right)+\frac{5C_Fn_f}{18}\bigg] \ ,
\end{align}
which agrees with our previous formula for $\gamma_{HH}=\Gamma_{S}$~\cite{Ji:2019ewn,Ji:2020ect,Ji:2021znw}.  In terms of above, one can express the following:
\begin{enumerate}
    \item The $\delta(1-z)$ term in the quark splitting function near $z=1$:
    \begin{align}
    P_{qq}(z)|_{z\rightarrow 1}\rightarrow \frac{2\Gamma_{\rm cusp}}{(1-z)^+}+2\gamma_{\phi} \delta(1-z) \ .
    \end{align}
The $\gamma_{\phi}$ (we use the notation in Ref.~\cite{Becher:2006mr}) is just the anomalous dimension for a on-shell quark attached to a light-like Wilson-line, and equals to
\begin{align}
\gamma_{\phi}=-\frac{\gamma_V}{2}-\frac{\gamma_s}{2} \ .
\end{align}
\item The constant term of the anomalous dimension $-2\gamma_{J}$~\cite{Becher:2006qw,Becher:2006mr} for the quark-jet function in DIS threshold factorization/quark beam function for TMD factorization of DY/SIDIS
\begin{align}
\gamma_{J}=-\gamma_{\phi}-\gamma_s=\frac{\gamma_V}{2}-\frac{\gamma_s}{2} \ .
\end{align}
Clearly, one has the relation $\gamma_{V}+\gamma_{\phi}-\gamma_{J}=0$.
\item The anomalous dimension $-\tilde \gamma_J$ for heavy-quark jet function~\cite{Jain:2008gb} is simply given in terms of $\gamma_{HL}$ and $\gamma_s$ as
\begin{align}
    \tilde \gamma_{J}=2\gamma_{HL}-2\gamma_s \ .
\end{align}
Notice that it remains the same for the space-like heavy quark jet function for quasi-PDF and the time-like heavy quark jet function. 
\item The UV anomalous dimension $\tilde \gamma_{H}$ for a conjugating pair of heavy-light Sudakov hard kernels~\cite{Ji:2019ewn,Ji:2020ect,Ji:2021znw, LPC:2022zci} can be expressed as
\begin{align}\label{eq:heavylighsuda}
\tilde \gamma_{H}=\gamma_{V}+2\gamma_{F}+2\gamma_{HL}-\gamma_s \ .
\end{align}
Explicitly, one has
\begin{align}
\tilde \gamma_H^{(1)}=&-\frac{C_F}{\pi} \ , \\ 
\tilde \gamma_H^{(2)}=&\frac{1188 \zeta_3-99 \pi ^2-1108}{432 \pi ^2}C_FC_A-\frac{36 \zeta_3+6-7 \pi ^2}{12 \pi ^2} C_F^2 \nonumber \\ 
&+\left( \frac{1}{24}+\frac{10}{27 \pi ^2}\right)C_Fn_f \ . \label{eq:tildegammaHresult}
\end{align}
Notice that the hard kernel for quasi-TMDPDF/quasi-LFWF factorization, as well as the threshold factorization for quasi-PDF are universal and is just the heavy-light Sudakov form factor, with the same anomalous dimension given in Eq.~(\ref{eq:heavylighsuda}).
\item The constant part of UV anomalous dimension of the standard TMD soft factor composed purely of light-like gauge-links is simply $-2\gamma_s$. 
\item The constant part of UV anomalous dimension for the TMD soft factor composed of one light-like Wilson-line staple and another space-like/time-like Wilson line staple~\cite{Ji:2021znw} is given by $-2\gamma_{HL}$.
\item Finally, notice that the constant term of UV anomalous dimension for quark TMDPDF/TMD fragmentation functions is simply given by $-\gamma_V\equiv-2\gamma_{H}$.
\end{enumerate}
To summarize, there exists a large number of relations between different anomalous dimensions ranging from threshold factorization to TMD factorization, very similar to the relations among different critical exponents 
for critical systems~\cite{mccoy_2015}. This should be expected, since high-energy limit of an asymptotically free QFT is indeed a critical system with logarithmic critical exponents.

\section{Space-like heavy quark jet function at NLO}\label{sec:NLOjet}
\begin{figure}[h!]
    \centering
    \includegraphics[height=3cm]{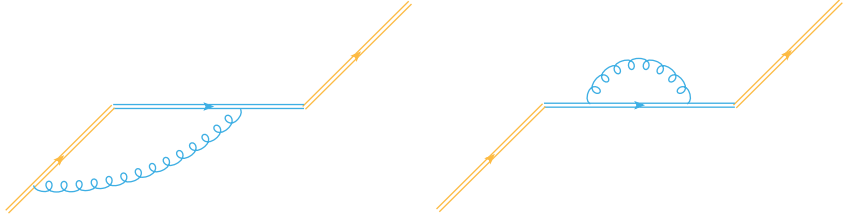}
    \caption{The one-loop vertex (left) and self-energy (right) diagrams for the jet function.}
    \label{fig:thresholdjet}
\end{figure}

In this appendix we present the one-loop calculation of the space-like jet function. There are two types of diagrams, the vertex diagram and the self-energy diagram, shown in Fig.~\ref{fig:thresholdjet}.
\subsubsection{Vertex-diagram}
The calculation is most easily done in coordinate space. For this purpose one needs the gluon propagator in $D$-dimensional coordinate space
\begin{align}
G_D^{\mu \nu}(x)=-g_{\mu \nu} \frac{\Gamma(\frac{D}{2}-1)\mu_0^{4-D}}{(4\pi)^{\frac{D}{2}}2^{2-D}}(-x^2+i0)^{1-\frac{D}{2}} \ .
\end{align}
Given these, the diagram with one-gluon exchange between $W_{n,-}$ and the $z$-direction link reads
\begin{align}
V_1(z>0)=(ig)^2C_F(-n\cdot n_z) \frac{\Gamma(\frac{D}{2}-1)\mu_0^{4-D}}{4\pi^{\frac{D}{2}}}\int_{-\infty}^{0} ds_1 \int_{0}^{z} ds_2 (-(n_zs_2-ns_1)^2+i0)^{1-\frac{D}{2}} \ .
\end{align}
Notice that $n\cdot n_z=-\frac{1}{\sqrt{2}}$, $-(n_zs_2-ns_1)^2=s_2^2-\sqrt{2}s_1s_2$.  In the region $z>0$, changing $s_1 \rightarrow -s_1$ one simply has
\begin{align}
V_1(z>0)=&-g^2C_F\frac{\Gamma(\frac{D}{2}-1)\mu_0^{4-D}}{4\sqrt{2}\pi^{\frac{D}{2}}}\int_{0}^{z} ds_2 \int_{0}^{\infty} ds_1(s_2^2+\sqrt{2}s_1s_2)^{1-\frac{D}{2}} \nonumber \\ 
&=\frac{g^2C_F\Gamma(\frac{D}{2}-1)(\mu_0 z)^{4-D}}{4\pi^{\frac{D}{2}}(D-4)^2} \ .
\end{align}
The other vertex diagram between $z$ and $W_{n,+}$ contributes equally.  In the region $z<0$, situation becomes tricky, one need to take care of the imaginary part. The contribution simply reads
\begin{align}
V_1(z<0)&=(ig)^2C_F(-n\cdot -n_z) \frac{\Gamma(\frac{D}{2}-1)\mu_0^{4-D}}{4\pi^{\frac{D}{2}}}\int_{-\infty}^{0} ds_1 \int_{0}^{|z|} ds_2 (-(n_zs_2+ns_1)^2+i0)^{1-\frac{D}{2}} \nonumber \\ 
&=g^2C_F\frac{\Gamma(\frac{D}{2}-1)\mu_0^{4-D}}{4\sqrt{2}}\int_{0}^{\infty}ds_1\int_{0}^{|z|} ds_2(s_2^2-\sqrt{2}s_2s_1+i0)^{1-\frac{D}{2}} \ .
\end{align}
This integral must be performed separately for $s_1>\frac{s_2}{\sqrt{2}}$ and $s_1<\frac{s_2}{\sqrt{2}}$, in the first region, a potential imaginary part can be generated. However, this contribution vanish in DR simply due to lacking of scale, as a result one simply has
\begin{align}
V_1(z<0)&=\frac{g^2C_F\Gamma(\frac{D}{2}-1)(\mu_0 |z|)^{4-D}}{4\pi^{\frac{D}{2}}(D-4)^2} \ .
\end{align}
In conclusion, the jet function depends only on $|z|$.

\subsubsection{Self-energy diagram}
Similarly, the self-energy diagram can be evaluated as
\begin{align}
S=&(ig)^2C_F(-n_z^2)\frac{\Gamma(\frac{D}{2}-1)\mu^{4-D}}{4\pi^{\frac{D}{2}}}\int_{0}^z ds_1 \int_{0}^{s_1} ds_2 (s_1-s_2)^{2-D} \nonumber \\ 
=&-\frac{g^2C_F \Gamma (\frac{D}{2}-1)(\mu_0 |z|)^{4-D}}{4\pi^{\frac{D}{2}}(3-D)(4-D)} \ .
\end{align}
The $3-D$ in the denominator corresponds to the linear divergence.

\section{Relation between timelike and spacelike jet functions}\label{sec:relation}
The timelike and spacelike jet functions, shown in Fig.~\ref{fig:timespacej}, are defined as follows
\begin{align}
\tilde J(t,D)=\langle \Omega|{\cal T}W_{n,+}(t n_t)W_t^{\dagger}(t n_t)W_t(0)W_{n,-}(0)|\Omega\rangle \ ,
\end{align}
\begin{align}
\tilde J(z,D)=\langle \Omega|{\cal T}W_{n,+}(zn_z)W_z^{\dagger}(zn_z)W_z(0)W_{n,-}(0)|\Omega\rangle \ , 
\end{align}
where the light-like vector $n=\frac{1}{\sqrt{2}}(1,1,0,0)$, the space-like vector $n_z=(0,1,0,0)$ and the timie-like vector $n_t=(1,0,0,0)$.
In this appendix, we show that they are related by simply replacing $t=-i |z|$ starting from the $t>0$ region, to all orders in perturbation theory
\begin{align}
\tilde J(t = -i |z|,D) = \tilde J(z,D).
\end{align}
The key issue in the proof is to show that one can do the Wick rotation in the loop integral without encountering any residue poles.

\subsection{A proof based on parametric-space representation}
First, we study the analyticity structure of the scalar field correlation function in coordinate space. Recall the Feynman propagator for scalar fields at tree level
\begin{align}
&G(x_1,x_2)=\langle \Omega | {\cal T} \phi(x_1) \phi(x_2) | \Omega \rangle = \int \frac{d^{D} k}{(2\pi)^D} \frac{i}{k^2+i 0} \exp\left[-i k (x_1-x_2)\right]\nonumber\\
&= \int \frac{d^{D} k}{(2\pi)^D} \int_{0}^{+\infty} d \alpha \exp \left[i \alpha (k^2+i 0) - i k (x_1-x_2)\right]\nonumber\\
&= \int_{0}^{+\infty} d \alpha \int \frac{d^{D} k}{(2\pi)^D} \exp \left[i \alpha \left(k-\frac{x_1-x_2}{2\alpha}\right)^2 - i \frac{(x_1-x_2)^2-i 0}{4 \alpha} \right]\nonumber\\
&= \int_{0}^{+\infty} d \alpha \frac{i}{(2\pi)^D}\frac{\pi^{D/2}}{(i \alpha)^{D/2}} \exp \left[- i \frac{(x_1-x_2)^2-i 0}{4 \alpha} \right]\nonumber\\
&= \frac{\Gamma  \left(\frac{D}{2}-1\right)}{4\pi^{D/2}} (-(x_1-x_2)^2+i 0)^{1-D/2},
\end{align}
where the $i0$ in the variable $-(x_1-x_2)^2+i0$ comes from the time ordering and is crucial in guaranteeing the exponential decay for $\alpha \rightarrow 0^{+}$. 

A general $m$-point scalar correlation function at arbitrary order in perturbation theory can be obtained through a number of contractions on a bunch of Feynman propagators. We start from the multiplication of $N/2$ Feynman propagators including $N$ different spacetime locations, 
\begin{align}
\prod_{i=1}^{N/2} G(x_{2 i-1},x_{2 i}),
\end{align}
which can be written in the following form,
\begin{align}\label{eq:gencorre}
I[{\cal P}, {\cal Q}, N, \{-(x_i-x_j)^2+i0\}] = \int_{0}^{+\infty} {\cal D} \alpha \, {\cal P}(\alpha) \exp\left[-\frac{i}{2} \sum_{i,j=1}^{N} {\cal Q}_{ij}(\alpha) (x_i-x_j)^2 \right],
\end{align}
where ${\cal Q} \geq 0$ and ${\cal Q}_{ij}={\cal Q}_{ji}$. $\alpha$ denotes a set of Schwinger parameters: $\alpha=\{\alpha_1,\alpha_2,...\}$ and $\int_{0}^{+\infty} {\cal D} \alpha = \int_{0}^{+\infty} \prod_{i} d \alpha_i$. This general correlation is a function of the relative spacetime intervals $\{-(x_i-x_j)^2+i0\}$ of the $N$ points, where $+i0$ is consistent with ${\cal Q} \geq 0$. Then we contract $M$ points ($M \leq N$),
\begin{align}
&\int d^D y \left[\int \prod_{j=1}^{M} d^D x_j \delta^{D}(x_j-y)\right] I[{\cal P}, {\cal Q}, N, \{-(x_i-x_j)^2+i0\}] \nonumber\\
&= \int_{0}^{+\infty} {\cal D} \alpha \, {\cal P}(\alpha) \int d^D y \exp\left[-i \sum_{i=M+1}^{N} \sum_{j=1}^{M} {\cal Q}_{ij}(\alpha) (x_i-y)^2 \right] \exp\left[-\frac{i}{2} \sum_{i,j=M+1}^{N} {\cal Q}_{ij}(\alpha) (x_i-x_j)^2 \right] \nonumber\\
&= \int_{0}^{+\infty} {\cal D} \alpha \, {\cal P}(\alpha) \int d^D y \exp\left[-i \sum_{i=M+1}^{N}{\cal \tilde{Q}}_{i}(\alpha) y^2 -i \frac{\sum_{i,j=M+1}^{N}(x_i-x_j)^2{\cal \tilde{Q}}_{i}(\alpha){\cal \tilde{Q}}_{j}(\alpha)}{2 \sum_{i=M+1}^{N} {\cal \tilde{Q}}_{i}(\alpha)} \right] \nonumber\\
&\exp\left[-\frac{i}{2} \sum_{i,j=M+1}^{N} {\cal Q}_{ij}(\alpha) (x_i-x_j)^2 \right]\nonumber\\
&=\int_{0}^{+\infty} {\cal D} \alpha \, {\cal P'}(\alpha) \exp\left[-\frac{i}{2} \sum_{i,j=M+1}^{N} {\cal Q'}_{ij}(\alpha) (x_i-x_j)^2 \right]\nonumber\\
&=I[{\cal P'}, {\cal Q'}, N-M, \{-(x_i-x_j)^2+i0\}],
\end{align}
where ${\cal \tilde{Q}}_{i}(\alpha)=\sum_{j=1}^{M} {\cal Q}_{ij}(\alpha)$, ${\cal Q'} \geq 0$ and ${\cal Q'}_{ij}={\cal Q'}_{ji}$. In principle ${\cal Q}'_{ij}$ can be expressed in terms of standard Graph polynomials~\cite{Bogner:2010kv}, but we will not need the explicit forms.  To summarize, the contraction keeps the form Eq.~(\ref{eq:gencorre}). The result after the contraction is a function of the relative spacetime intervals $\{-(x_i-x_j)^2+i0\}$ of the remaining $N-M$ points, where $+i0$ is consistent with ${\cal Q'} \geq 0$. One can keep doing the contractions over and over again until obtain the desired $m$-point correlation function, which is a function of the relative spacetime intervals $\{-(x_i-x_j)^2+i0\}$ of the $m$ points. 

We can introduce Feynman parameters in Eq.~(\ref{eq:gencorre}): $\alpha_i = \xi_i \rho$. $\xi_i$ is the dimensionless Feynman parameter and $\rho$ has the dimension.  
For $m$ Schwinger parameters, the integral measurement becomes $$\int_{0}^{+\infty} \prod_{i=1}^{m} d \alpha_i = \int_{0}^{1}\prod_{i=1}^{m} d \xi_i \int_{0}^{+\infty} d \rho \, \rho^{m-1} \delta\left(1-\sum_{i=1}^{m}\xi_i\right)$$
And Eq.~(\ref{eq:gencorre}) for $m$ Schwinger parameters becomes
\begin{align}\label{eq:gencorre2}
&I[{\cal P}, {\cal Q}, N, \{-(x_i-x_j)^2+i0\}] \nonumber\\
&= \int_{0}^{1}\prod_{i=1}^{m} d \xi_i \int_{0}^{+\infty} d \rho \, \rho^{m-1} \delta\left(1-\sum_{i=1}^{m}\xi_i\right) \, {\cal P}(\xi \rho) \exp\left[-\frac{i}{2} \sum_{i,j=1}^{N} {\cal Q}_{ij}(\xi \rho) (x_i-x_j)^2 \right]
\end{align}

Then, an $m$-point gluon correlation function is defined as 
\begin{align}
G_{\mu_1...\mu_m}^{a_1 ... a_m}(x_1,..,x_m) = \langle \Omega |{\cal T} A(x_1)_{\mu_1}^{a_1} ... A(x_m)_{\mu_m}^{a_m} | \Omega \rangle,
\end{align}
where $\mu_1...\mu_m$ are Lorentz indices for the gluon fields, $a_1 ... a_m$ are color indices and $x_1,..,x_m$ are the spacetime locations. This is a general correlation function including both fully connected and separately connected parts. But vacuum bubbles are not included. The color, spin and scalar structures can be factorized out
\begin{align}
G_{\mu_1...\mu_m}^{a_1 ... a_m}(x_1,..,x_m) = \sum_{i} C_{i}^{a_1 ... a_m} \Gamma^{i}_{\mu_1...\mu_m}(x_1,..,x_m) G_{i}(\{-(x_j-x_k)^2+i0|j\neq k\}),
\end{align}
where $C_{i}^{a_1 ... a_m}$ denotes the color structure, which is independent of $x_j$. $\Gamma^{i}_{\mu_1...\mu_m}(x_1,..,x_m)$ is the spin structure, which can contain the metric tensor $g^{\mu \nu}$ and the spacetime vectors $x_j^{\mu}$. We absorb all the residue poles in the scalar structure $G_{i}$, which is a function of all the relative spacetime intervals $\{-(x_j-x_k)^2+i0|j\neq k\}$ as we have argued above. Here $i0$ comes from the time ordering, which is crucial in our proof. $\sum_{i}$ means summing up all the possible combinations of color structure, spin structure and scalar structure. 

Then the timelike jet functions can be written as the gluon correlation function attached to the Wilson links
\begin{align}
\tilde J(t,D)=&\sum_{m=0}^{+\infty} \sum_{m_1,m_2} \frac{1}{m_1! m_2! (m-m_1-m_2)!} {\cal P} \int^{t n_t + \infty n}_{t n_t} d x_1^{\mu_1} ... d x_{m_1}^{\mu_{m_1}} \int^{t n_t}_{0} d x_{m_1 + 1}^{\mu_{m_1+1}} ... d x_{m_1 + m_2}^{\mu_{m_1+m_2}} \nonumber\\ 
&\int^{0}_{-\infty n} d x_{m_1 + m_2 + 1}^{\mu_{m_1+m_2+1}} ... d x_{m}^{\mu_m} (-i g t_{a_1})...(-i g t_{a_{m}}) G_{\mu_1...\mu_m}^{a_1 ... a_m}(x_1,..,x_m),
\end{align}
where $m_1$ gluons are attached to $W_{n,+}(t n_t)$, $m_2$ gluons are attached to $W_t^{\dagger}(t n_t)W_t(0)$, and $m-m_1-m_2$ gluons are attached to $W_{n,-}(0)$. $\sum_{m_1,m_2}$ means summing up all the possibilities for distributing $m$ into $m_1$, $m_2$ and $m-m_1-m_2$. ${\cal P}$ is the path ordering operator along the positive direction of the gauge links, which is $-\infty n \rightarrow 0 \rightarrow t n_t \rightarrow t n_t + \infty n$, and one needs to permutate the color matrices $(-i g t_{a_1})...(-i g t_{a_{m}})$ according to the path ordering. 

We parametrize the spacetime coordinates as
\begin{align}
x_j(S,tn_t,n)= 
\left\{
    \begin{array}{lr}
        s_j n + t n_t, & \text{for } j=1,...,m_1\\
        s_j n_t, & \text{for } j=m_1+1,...,m_1+m_2\\
        s_j n, & \text{for } j=m_1+m_2+1,...,m 
    \end{array}
\right\}
\end{align}
To simplify the notation, we introduce $S^{n,+}=\{s_1,..,s_{m_1}\}$, $S^{t}=\{s_{m_1 + 1},...,s_{m_1 + m_2}\}$ and $S^{n,-}=\{s_{m_1 + m_2 + 1},...,s_{m}\}$. And we will use $S^{n,+}_j$ (for $j=1,...,m_1$), $S^{t}_j$ (for $j=m_1+1,...,m_1+m_2$) and $S^{n,-}_j$ (for $j=m_1+m_2+1,...,m$) to denote the elements in these sets. We introduce $C_{a_1...a_m}=(-i g t_{a_1})...(-i g t_{a_{m}})$ for color structures. We introduce $L^{\mu_1 ... \mu_{m_1}}_{n,+}=n^{\mu_1}..n^{\mu_{m_1}}$, $L^{\mu_{m_1+1} ... \mu_{m_1+m_2}}_{t}=n_t^{\mu_{m_1+1}}..n_t^{\mu_{m_1+m_2}}$ and $L^{\mu_{m_1+m_2+1} ... \mu_{m}}_{n,-}=n^{\mu_{m_1+m_2+1}}..n^{\mu_{m}}$ for spin structures. The timelike jet function becomes
\begin{align}
\tilde J( t,D)&=\sum_{m=0}^{+\infty} \sum_{m_1,m_2} \frac{1}{m_1! m_2! (m-m_1-m_2)!} {\cal P} \int^{+ \infty}_{0} {\cal D} S^{n,+} \int^{t}_{0} {\cal D} S^{t} 
\int^{0}_{-\infty} {\cal D} S^{n,-} \nonumber\\ 
&C_{a_1...a_m} L^{\mu_1 ... \mu_{m_1}}_{n,+} L^{\mu_{m_1+1} ... \mu_{m_1+m_2}}_{t} L^{\mu_{m_1+m_2+1} ... \mu_{m}}_{n,-} G_{\mu_1...\mu_m}^{a_1 ... a_m}(x_1,..,x_m) \nonumber\\ 
&=\sum_{m=0}^{+\infty} \sum_{m_1,m_2} \sum_{i} \frac{1}{m_1! m_2! (m-m_1-m_2)!} {\cal P} \int^{+ \infty}_{0} {\cal D} S^{n,+} \int^{t}_{0} {\cal D} S^{t} 
\int^{0}_{-\infty} {\cal D} S^{n,-} \nonumber\\ 
&C_{a_1...a_m} C_{i}^{a_1 ... a_m} L^{\mu_1 ... \mu_{m_1}}_{n,+} L^{\mu_{m_1+1} ... \mu_{m_1+m_2}}_{t} L^{\mu_{m_1+m_2+1} ... \mu_{m}}_{n,-} \Gamma^{i}_{\mu_1...\mu_m}(S^{n,+} n + t n_t,S^{t} n_t,S^{n,-} n) \nonumber\\ 
&G_{i}\left(\{-(S^{n,+}_j-S^{n,+}_k)^2 n^2+i0|j\neq k\},\{-(S^{t}_j-S^{t}_k)^2 n_t^2+i0|j\neq k\}, \right. \nonumber\\
& \left. \{-(S^{n,-}_j-S^{n,-}_k)^2 n^2+i0|j\neq k\},
\{-(S^{n,+}_j n + t n_t - S^{t}_k n_t)^2+i0|j\neq k\}, \right. \nonumber\\
& \left. \{-(S^{t}_j n_t - S^{n,-}_k n)^2+i0|j\neq k\},\{-(S^{n,+}_j n + t n_t - S^{n,-}_k n)^2+i0|j\neq k\} \right),
\end{align}
where we classify the relative spacetime intervals (which are variables of the scalar structure $G_i$) into six categories according to whether the spacetime points are on the $n+$, $t$ or $n-$ Wilson links. And only four categories are left since $n^2=0$,
\begin{align}\label{timelikeJetf}
\tilde J(t,D)&=\sum_{m=0}^{+\infty} \sum_{m_1,m_2} \sum_{i} \frac{1}{m_1! m_2! (m-m_1-m_2)!} {\cal P} \int^{+ \infty}_{0} {\cal D} S^{n,+} \int^{t}_{0} {\cal D} S^{t} 
\int^{+\infty}_{0} {\cal D} S^{n,-} \nonumber\\
&\left(C \cdot C_{i}\right) \left(L_{n,+}\, L_{t} \, L_{n,-} \, \Gamma^{i}(S^{n,+} n + t n_t,S^{t} n_t,-S^{n,-} n)\right) \nonumber\\ 
&G_{i}\left(0,\{-(S^{t}_j-S^{t}_k)^2+i0\}, 0,
\{-(t - S^{t}_k)^2-\sqrt{2} S^{n,+}_j (t-S^{t}_k)+i0\}, \right. \nonumber\\
& \left. \{-(S^{t}_j)^2 - \sqrt{2} S^{t}_j S^{n,-}_k+i0\},\{-t^2-\sqrt{2} t (S^{n,+}_j + S^{n,-}_k)+i0\} \right),
\end{align}
where we omit the color indices, spin indices and $j\neq k$ in the spacetime intervals for simplicity. Following the same logic, one can express the space-like Jet function with the same gluon correlation function
\begin{align}\label{spacelikeJetf}
\tilde J(z,D)&=\sum_{m=0}^{+\infty} \sum_{m_1,m_2} \sum_{i} \frac{1}{m_1! m_2! (m-m_1-m_2)!} {\cal P} \int^{+ \infty}_{0} {\cal D} S^{n,+} \int^{z}_{0} {\cal D} S^{z} 
\int^{+\infty}_{0} {\cal D} S^{n,-} \nonumber\\
&\left(C \cdot C_{i}\right) \left(L_{n,+}\, L_{z} \, L_{n,-} \, \Gamma^{i}(S^{n,+} n + z n_z,S^{z} n_z,-S^{n,-} n)\right) \nonumber\\ 
&G_{i}\left(0,\{(S^{z}_j-S^{z}_k)^2+i0\}, 0,
\{(z - S^{z}_k)^2+\sqrt{2} S^{n,+}_j (z-S^{z}_k)+i0\}, \right. \nonumber\\
& \left. \{(S^{z}_j)^2 + \sqrt{2} S^{z}_j S^{n,-}_k+i0\},\{z^2+\sqrt{2} z (S^{n,+}_j + S^{n,-}_k)+i0\} \right),
\end{align}
where $L^{\mu_{m_1+1} ... \mu_{m_1+m_2}}_{z}=n_z^{\mu_{m_1+1}}...n_z^{\mu_{m_1+m_2}}$.

In the time-like Jet function, one can introduce the dimensionless Feynman parameters $\Xi^{n,+}=\{\xi_1,...,\xi_{m_1}\}$, $\Xi^{t}=\{\xi_{m_1+1},...,\xi_{m_1+m_2}\}$ and $\Xi^{n,-}=\{\xi_{m_1+m_2+1},...,\xi_{m}\}$. Then one can write the spacetime intervals with Feynman parameters, $S^{n,+}_{j}=\xi_j s$, $S^{t}_j=\xi_j t$ and $S^{n,-}_j=\xi_j s$, where $s$ has the length dimension and $0 \leq \xi_j \leq 1$. The time-like Jet function becomes
\begin{align}
\tilde J(t,D)&=\sum_{m=0}^{+\infty} \sum_{m_1,m_2} \sum_{i} \frac{1}{m_1! m_2! (m-m_1-m_2)!} {\cal P} \int^{1}_{0} {\cal D} \Xi^{n,+} 
\int^{1}_{0} {\cal D} \Xi^{t}
\int^{1}_{0} {\cal D} \Xi^{n,-}
{\cal J}(\Xi^{n,+},\Xi^{t},\Xi^{n,-})
\nonumber\\
&\int^{+\infty}_{0} ds \, t^{m_2} s^{m-m_2-1} \left(C \cdot C_{i}\right) \left(L_{n,+}\, L_{t} \, L_{n,-} \, \Gamma^{i}(S^{n,+} n + t n_t,S^{t} n_t,-S^{n,-} n)\right) \nonumber\\ 
&G_{i}\left(0,
\{-(\xi_j-\xi_k)^2 t^2+i0\}, 
0,
\{-(1 - \xi_k)^2 t^2 -\sqrt{2} \xi_j (1-\xi_k) t s+i0\}, \right. \nonumber\\
& \left. \{-(\xi_j)^2 t^2 - \sqrt{2} \xi_j \xi_k t s+i0\},
\{-t^2-\sqrt{2} (\xi_j + \xi_k ) t s+i0\} \right),
\end{align}
where ${\cal J}(\Xi^{n,+},\Xi^{t},\Xi^{n,-})=\delta\left(\sum_{i=1}^{m_1}\xi_i+\sum_{i=m_1+m_2+1}^{m}\xi_i-1\right)$ is the dimensionless part of the Jacobian. On the complex plane of $s$, the residue poles only exist in Quadrant II (One can check all the variables of $G_i$ including $\{-(\xi_j-\xi_k)^2 t^2+i0\}$, $\{-(1 - \xi_k)^2 t^2 -\sqrt{2} \xi_j (1-\xi_k) t s+i0\}$, $\{-(\xi_j)^2 t^2 - \sqrt{2} \xi_j \xi_k t s+i0\}$ and $\{-t^2-\sqrt{2} (\xi_j + \xi_k ) t s+i0\}$). So one can do the Wick rotation $s \rightarrow -i s$ without encountering any poles. In parametric representation of $G_i$, this Wick rotation keeps the exponential decay, see Eq.~(\ref{eq:gencorre2}). Then one can simply replace $t \rightarrow -i z$ since it is an analytical function of $t$. The time-like Jet function becomes 
\begin{align}
\tilde J(t,D)&=\sum_{m=0}^{+\infty} \sum_{m_1,m_2} \sum_{i} \frac{1}{m_1! m_2! (m-m_1-m_2)!} {\cal P} \int^{1}_{0} {\cal D} \Xi^{n,+} 
\int^{1}_{0} {\cal D} \Xi^{t}
\int^{1}_{0} {\cal D} \Xi^{n,-}
{\cal J}(\Xi^{n,+},\Xi^{t},\Xi^{n,-})
\nonumber\\
&\int^{+\infty}_{0} ds \, (-i)^{m} z^{m_2} s^{m-m_2-1} \left(C \cdot C_{i}\right) \left(L_{n,+}\, L_{t} \, L_{n,-} \, \Gamma^{i}(-i S^{n,+} n-i z n_t,-i S^{t} n_t,i S^{n,-} n)\right) \nonumber\\ 
&G_{i}\left(0,
\{(\xi_j-\xi_k)^2 z^2+i0\}, 
0,
\{(1 - \xi_k)^2 z^2 + \sqrt{2} \xi_j (1-\xi_k) z s+i0\}, \right. \nonumber\\
& \left. \{(\xi_j)^2 z^2 + \sqrt{2} \xi_j \xi_k z s+i0\},
\{z^2+\sqrt{2} (\xi_j + \xi_k ) z s+i0\} \right)
\end{align}
Then one can rewrite the integral with the $S^{n,+}_{j}$, $S^{t}_j$, $S^{n,-}_j$ parameters. The final result is the same as simply replacing $S_{j} \rightarrow -i S_{j}$ and $t \rightarrow -i z$ in Eq.~(\ref{timelikeJetf}),
\begin{align}\label{timelikeJetfWR}
\tilde J(t,D)&=\sum_{m=0}^{+\infty} \sum_{m_1,m_2} \sum_{i} \frac{1}{m_1! m_2! (m-m_1-m_2)!} {\cal P} \int^{+ \infty}_{0} {\cal D} S^{n,+} \int^{z}_{0} {\cal D} S^{t} 
\int^{+\infty}_{0} {\cal D} S^{n,-} \nonumber\\
&\left(C \cdot C_{i}\right) (-i)^m \left(L_{n,+}\, L_{t} \, L_{n,-} \, \Gamma^{i}(-i S^{n,+} n - i z n_t,-i S^{t} n_t,i S^{n,-} n)\right) \nonumber\\ 
&G_{i}\left(0,\{(S^{t}_j-S^{t}_k)^2+i0\}, 0,
\{(z - S^{t}_k)^2+\sqrt{2} S^{n,+}_j (z-S^{t}_k)+i0\}, \right. \nonumber\\
& \left. \{(S^{t}_j)^2 + \sqrt{2} S^{t}_j S^{n,-}_k+i0\},\{z^2 + \sqrt{2} z (S^{n,+}_j + S^{n,-}_k)+i0\} \right),
\end{align}
where the scalar structure $G_i$ is exactly the same as that in Eq.~(\ref{spacelikeJetf}). 

Now Let's study the spin structures. The general form of the spin structure in the time-like Jet function Eq.~(\ref{timelikeJetf}) is
\begin{align}
&L_{n,+}\, L_{t} \, L_{n,-} \, \Gamma^{i}(S^{n,+} n + t n_t,S^{t} n_t,-S^{n,-} n) \nonumber\\
&= \left(\prod^{k_1}_{j=1} n \cdot (S_{j}^{n,+} n + t n_t)\right) 
\left(\prod^{k_2}_{j=1} n_t \cdot S_j^{t} n_t\right) 
\left(\prod^{k_3}_{j=1} n_t \cdot (S_{j}^{n,+} n + t n_t)\right)  \nonumber\\
&\left(\prod^{k_4}_{j=1} n \cdot S_j^{t} n_t\right) 
\left(\prod^{k_5}_{j=1} -S^{n,-} n_t \cdot n\right)
\left(\prod^{k_6}_{j=1} n_t \cdot n\right) \left(\prod^{k_7}_{j=1} n_t \cdot n_t\right) \nonumber\\
&= \left(\frac{\sqrt{2}}{2}t \right)^{k_1} 
\left(\prod^{k_2}_{j=1} S_j^{t}\right) 
\left(\prod^{k_3}_{j=1} \frac{\sqrt{2}}{2} S_{j}^{n,+} + t\right) 
\left(\prod^{k_4}_{j=1} \frac{\sqrt{2}}{2} S_j^{t} \right) 
\left(\prod^{k_5}_{j=1} -\frac{\sqrt{2}}{2} S^{n,-}\right)
\left(\frac{\sqrt{2}}{2}\right)^{k_6},
\end{align}
where $k_1+k_2+k_3+k_4+k_5+2 k_6+2 k_7=m$. Here we don't consider the contractions between $S^{n,+} n + t n_t,S^{t} n_t,-S^{n,-} n$ since we absorb that in the scalar structure. After the Wick rotation, the spin structure multiplied by $(-i)^m$ in Eq.~(\ref{timelikeJetfWR}) is 
\begin{align}
&(-i)^m L_{n,+}\, L_{t} \, L_{n,-} \, \Gamma^{i}(-i S^{n,+} n - i z n_t,-i S^{t} n_t,i S^{n,-} n) \nonumber\\
&= (-i)^m \left(-i \frac{\sqrt{2}}{2}z \right)^{k_1} 
\left(\prod^{k_2}_{j=1} -i S_j^{t}\right) 
\left(\prod^{k_3}_{j=1} -i \frac{\sqrt{2}}{2} S_{j}^{n,+} - i z\right) \left(\prod^{k_4}_{j=1} -i \frac{\sqrt{2}}{2} S_j^{t} \right) 
\left(\prod^{k_5}_{j=1} i \frac{\sqrt{2}}{2} S^{n,-}\right) \left(\frac{\sqrt{2}}{2}\right)^{k_6} \nonumber\\
&= (-i)^m i^{m-2k_6-2k_7} \left(- \frac{\sqrt{2}}{2}z \right)^{k_1} 
\left(\prod^{k_2}_{j=1} - S_j^{t}\right) 
\left(\prod^{k_3}_{j=1} - \frac{\sqrt{2}}{2} S_{j}^{n,+} -  z\right) \left(\prod^{k_4}_{j=1} - \frac{\sqrt{2}}{2} S_j^{t} \right) 
\left(\prod^{k_5}_{j=1} \frac{\sqrt{2}}{2} S^{n,-}\right) \left(\frac{\sqrt{2}}{2}\right)^{k_6} \nonumber\\
&= (-1)^{k_6+k_7} \left(- \frac{\sqrt{2}}{2}z \right)^{k_1} 
\left(\prod^{k_2}_{j=1} - S_j^{t}\right) 
\left(\prod^{k_3}_{j=1} - \frac{\sqrt{2}}{2} S_{j}^{n,+} -  z\right) \left(\prod^{k_4}_{j=1} - \frac{\sqrt{2}}{2} S_j^{t} \right) 
\left(\prod^{k_5}_{j=1} \frac{\sqrt{2}}{2} S^{n,-}\right) \left(\frac{\sqrt{2}}{2}\right)^{k_6},
\end{align}
which is exactly the same as the general form of the spin structure in the space-like Jet function Eq.~(\ref{spacelikeJetf})
\begin{align}
&L_{n,+}\, L_{z} \, L_{n,-} \, \Gamma^{i}(S^{n,+} n + z n_z,S^{z} n_z,-S^{n,-} n) \nonumber\\
&= \left(-\frac{\sqrt{2}}{2}z \right)^{k_1} 
\left(\prod^{k_2}_{j=1} -S_j^{z}\right) 
\left(\prod^{k_3}_{j=1} -\frac{\sqrt{2}}{2} S_{j}^{n,+} -z \right) 
\left(\prod^{k_4}_{j=1} -\frac{\sqrt{2}}{2} S_j^{z} \right) 
\left(\prod^{k_5}_{j=1} \frac{\sqrt{2}}{2} S^{n,-}\right)
\left(-\frac{\sqrt{2}}{2} \right)^{k_6}
\left(-1 \right)^{k_7}
\end{align}

So Eq.~(\ref{timelikeJetfWR}) is the same as Eq.~(\ref{spacelikeJetf}). Thus the space-like Jet function is equal to the time-like Jet function with $t \rightarrow -i z$.

\subsection{A proof based on general property of Wightman functions}
The above proof can be generalized to the case where a non-perturbative Wightman function attached to the Wilson links through perturbative vertices. In Eq.~(\ref{timelikeJetf}), requiring the time ordering, one can rewrite $G_i$ in terms of a Wightman function ${\cal W}$,
\begin{align}
\tilde J(t,D)&=\sum_{m=0}^{+\infty} \sum_{m_1,m_2} \sum_{i} \int^{+ \infty}_{0} d s_1 \int^{s_1}_{0} d s_2 ... \int^{s_{m_1-1}}_{0} d s_{m_1} 
\int^{t}_{0} d s_{m_1+1} ... \int^{s_{m_1+m_2-1}}_{0} d s_{m_1+m_2} \nonumber\\
&\int^{0}_{-\infty} d s_{m_1+m_2+1} ... \int^{s_{m-1}}_{-\infty} d s_{m}
\left(C \cdot C_{i}\right)L_{n,+}\, L_{t} \, L_{n,-}\nonumber \\ 
&\cdot{\cal W}\bigg((x_{i}(S,tn_t,n)-x_{i+1}(S,tn_t,n))(1-i0)\bigg) \ ,
\end{align}
where the Wightman function ${\cal W}$ contains the Lorentz vector indices contracted with $L_{n,+}\, L_{t} \, L_{n,-} \,$ and we just omit them here for simplicity.  The Wightman function depends on the consecutive increments according to the natural time ordering. More explicitly, one has
\begin{align}
&\bigg(x_{i}-x_{i+1}\bigg)(S,tn_t,n)=(s_i-s_{i+1})n \ , \ \ 1\le i\le m_1-1 \ , \\
&\bigg(x_{m_1}-x_{m_1+1}\bigg)(S,tn_t,n)=s_{m_1}n+(1-s_{m_{1}+1})tn_t \ , \\
&\bigg(x_{i}-x_{i+1}\bigg)(S,tn_t,n)=(s_i-s_{i+1})tn_t\ ,  \ \ m_1+1\le i\le m_1+m_2-1 \ , \\
&\bigg(x_{m_1+m_2}-x_{m_1+m_2+1}\bigg)(S,tn_t,n)=s_{m_1+m_2}tn_t-s_{m_1+m_2+1}n\ , \\
&\bigg(x_{i}-x_{i+1}\bigg)(S,tn_t,n)=(s_i-s_{i+1})n \ , \ m_1+m_2+1\le i\le m-1 \ .
\end{align}
The $i0$ choice guarantees that the consecutive increments in this Wightman function are all within the forward lightcone direction, namely, $x_{i}-x_{i+1} \in R^4-iV_+ $, the natural analyticity domain. So one can do the analytical continuation $s_j \rightarrow -i s_j$ and $t \rightarrow -i z$ to obtain
\begin{align}\label{eq:tlJetAC}
&\tilde J(t=-i|z|,D)\nonumber\\
&=\sum_{m=0}^{+\infty} \sum_{m_1,m_2} \sum_{i} \int^{+ \infty}_{0} d s_1 \int^{s_1}_{0} d s_2 ... \int^{s_{m_1-1}}_{0} d s_{m_1} 
\int^{z}_{0} d s_{m_1+1} ... \int^{s_{m_1+m_2-1}}_{0} d s_{m_1+m_2} \nonumber\\
&\int^{0}_{-\infty} d s_{m_1+m_2+1} ... \int^{s_{m-1}}_{-\infty} d s_{m}
\left(C \cdot C_{i}\right) (-i)^{m}L_{n,+}\, L_{t} \, L_{n,-} \nonumber \\ 
&\cdot
{\cal W}\bigg(-ix_i(S,zn_t,n)+ix_{i+1}(S,zn_t,n)\bigg).
\end{align}
According to the general analyticity property of the Wightman function~\cite{Streater:1989vi}, $L_{n,+}\, L_{t} \, L_{n,-} \cdot {\cal W}$ is invariant under the complex Lorentz transformation ${\Lambda}(tn_t+zn_z+\vec{x}_{\perp})=itn_z+izn_t+{\vec x}_{\perp}$ which transforms $n_t \rightarrow -in_z$ and $n \rightarrow -in$. The inverse transformation is $\Lambda^{-1}=-\Lambda$. The vector field transforms as $UA^{\mu}(x)U^{\dagger}=\Lambda^{\mu}_{\nu}A^{\nu}(\Lambda^{-1}x)$, which implies that ${\cal W}^{\mu_1...\mu_n}(x_1,..x_n)=(\Lambda^{-1})^{\mu_1}_{\nu_1}...(\Lambda^{-1})^{\mu_n}_{\nu_n}{\cal W}^{\nu_1....\nu_n}(\Lambda x_1,..\Lambda x_n)$ in the analyticity domain under complex Lorentz transform $\Lambda$ with ${\rm det} \  \Lambda=1$.

So we can obtain the following identity
\begin{align}
&L_{n,+}\, L_{z} \, L_{n,-} \cdot \cdot {\cal W}\bigg(x_i(S,zn_z,n)-x_{i+1}(S,zn_t,n)\bigg)
\nonumber \\
&= (-i)^m L_{n,+}\, L_{t} \, L_{n,-} \cdot {\cal W}\bigg(-ix_i(S,zn_t,n)+ix_{i+1}(S,zn_t,n)\bigg) \ ,
\end{align}
where $x_i(S,zn_z,n)$ is defined exactly as $x_i(S,tn_t,n)$ with $tn_t$ replaced by $zn_z$. Here we used the fact that $-i\Lambda x_i(S,zn_t,n)=x_i(S,zn_z,n)$, $-i(n_t)_{\mu}(\Lambda^{-1})^{\mu}_{\nu}v^{\nu}=-v^z=(n_z)_{\mu}v^{\mu}$ and $-i(n_z)_{\mu}(\Lambda^{-1})^{\mu}_{\nu}v^{\nu}=v^t=(n_t)_{\mu}v^{\mu}$. Plug the above identity into Eq.~(\ref{eq:tlJetAC}), we obtain
\begin{align}
&\tilde J(t=-i |z|,D)\nonumber\\
&=\sum_{m=0}^{+\infty} \sum_{m_1,m_2} \sum_{i} \int^{+ \infty}_{0} d s_1 \int^{s_1}_{0} d s_2 ... \int^{s_{m_1-1}}_{0} d s_{m_1} 
\int^{z}_{0} d s_{m_1+1} ... \int^{s_{m_1+m_2-1}}_{0} d s_{m_1+m_2} \nonumber\\
&\int^{0}_{-\infty} d s_{m_1+m_2+1} ... \int^{s_{m-1}}_{-\infty} d s_{m}
\left(C \cdot C_{i}\right) L_{n,+}\, L_{z} \, L_{n,-} \cdot
{\cal W}\bigg(x_i(S,zn_z,n)-x_{i+1}(S,zn_z,n)\bigg),
\end{align}
which is the same as the spacelike jet function (Eq.~(\ref{spacelikeJetf})), written with the Wightman function,
\begin{align}
&\tilde J(|z|,D)\nonumber\\
&=\sum_{m=0}^{+\infty} \sum_{m_1,m_2} \sum_{i} \int^{+ \infty}_{0} d s_1 \int^{s_1}_{0} d s_2 ... \int^{s_{m_1-1}}_{0} d s_{m_1} 
\int^{z}_{0} d s_{m_1+1} ... \int^{s_{m_1+m_2-1}}_{0} d s_{m_1+m_2} \nonumber\\
&\int^{0}_{-\infty} d s_{m_1+m_2+1} ... \int^{s_{m-1}}_{-\infty} d s_{m}
\left(C \cdot C_{i}\right) L_{n,+}\, L_{z} \, L_{n,-} \cdot
{\cal W}\bigg(x_i(S,zn_z,n)-x_{i+1}(S,zn_z,n)\bigg) \ .
\end{align}
This shows that the time-like jet function simply relates to the space--like version through the analytic transform $t\rightarrow -iz$ when $t>0$ and $z>0$.

\section{Fourier transform of logarithms}\label{sec:fourier}
In this appendix we provide a pedagogical introduction to Fourier transforms of logarithm, emphasizing the singular contribution at $x=0$. We first introduce for $\varphi \in {\cal S}(R)$ (the Schwartz class), 
\begin{align}\label{eq:deflog}
\langle |z|^{\alpha},\varphi(z)\rangle\equiv \int_{-\infty}^{\infty} |z|^{\alpha} \varphi(z)dz \ .
\end{align}
Clearly, this defines for any $-1<\alpha<\infty$ a tempered distribution acting on smooth functions  $\varphi \in {\cal S}(R)$ with fast decrease at large $z$. Furthermore, it is easy to see that it is infinitely smooth near $\alpha=0$, with
\begin{align}
\frac{d^n}{d^n\alpha}\langle |z|^{\alpha},\varphi(z)\rangle_{\alpha=0}=\int_{-\infty}^{\infty} \ln^n |z| \varphi(z)dz \ .
\end{align}
We would like to express the distribution Eq.~(\ref{eq:deflog}) in terms of $\hat \varphi(x)$, the Fourier transform of $\varphi(z)$, through absolute convergent integral representations. The key point is, for $0<\alpha<-1$, Fourier transform of $|z|^{\alpha}$ is of the form $|x|^{-1+|\alpha|}$, which is integrable at $|x|=0$ and requires no subtraction at all. Therefore one needs to partial integrate to decrease the power of $|z|^{\alpha}$ down to $|z|^{\alpha-1}$. This can be performed as below
\begin{align}
&\int_{-\infty}^{\infty} dz |z|^{\alpha}\int_{-\infty}^{\infty}e^{ixz}\hat \varphi(x) dx=2\int_{0}^{\infty} dz z^{\alpha-1}\int_{-\infty}^{\infty}z\cos xz\hat \varphi(x) dx\nonumber \\ 
&=-2\int_{0}^{\infty} dz z^{\alpha-1}\int_{-\infty}^{\infty}\sin xz\hat \varphi'(x) dx=-2\Gamma(\alpha)\sin \frac{\alpha \pi}{2}\int_{-\infty}^{\infty} dx \frac{\hat \varphi'(x) {\rm sign}(x)}{|x|^{\alpha}} \ .
\end{align}
Clearly, for $0<\alpha<1$ this integral is convergent absolutely near $x=0$. Now, we simply partial integrate 
\begin{align}
&\int_{-\infty}^{\infty} dx \frac{\hat \varphi'(x) {\rm sign}(x)}{|x|^{\alpha}} \nonumber \\ &=\int_{0}^1 dx \frac{(\hat \varphi(x)-\hat \varphi(0))'}{|x|^{\alpha}}+\int_{1}^{\infty} dx\frac{\hat \varphi'(x)}{|x|^{\alpha}}-\int_{-1}^0 dx \frac{(\hat \varphi(x)-\hat \varphi(0))'}{|x|^{\alpha}}-\int_{-\infty}^{-1} dx\frac{\hat \varphi'(x)}{|x|^{\alpha}} \nonumber \\ 
&=\alpha\int_{-1}^{1}dx\frac{\hat \varphi(x)-\hat \varphi(0)}{|x|^{1+\alpha}}+\alpha\int_{|x|>1}dx\frac{\hat \varphi(x)}{|x|^{1+\alpha}}-2\hat \varphi(0) \ .
\end{align}
Therefore, one finally has the master formula
\begin{align}\label{eq:FTeven}
\int_{-\infty}^{\infty} |z|^{\alpha} \varphi(z)dz=2\Gamma(\alpha)\sin \frac{\pi \alpha}{2}\bigg(2\hat \varphi(0)-\alpha\int_{-1}^{1}dx\frac{\hat \varphi(x)-\hat \varphi(0)}{|x|^{1+\alpha}}-\alpha\int_{|x|>1}dx\frac{\hat \varphi(x)}{|x|^{1+\alpha}}\bigg) \ .
\end{align}
Now, all the integral converges absolutely and defines a smooth functions of $\alpha$ in a neighborhood of $\alpha=0$, therefore by taking derivative at $\alpha=0$ and equating two sides of the equation, one generates Fourier transform of all the $\ln^n |z|$. For example
\begin{align}
&{\cal F}(\ln |z|)=-\gamma_E\delta(x)-{\cal P}\frac{1}{2|x|} \ , \\
&{\cal F}(\ln^2 |z|)=\bigg(\gamma_E^2+\frac{\pi^2}{12}\bigg)\delta(x)+{\cal P}\frac{\ln e^{\gamma_E}|x|}{|x|} \ , \\ 
&{\cal F}(\ln^3 |z|)=-\bigg(\gamma_E^3+\frac{1}{4}\gamma_E\pi^2+2\zeta_3\bigg)\delta(x)-{\cal P}\frac{12\ln^2 e^{\gamma_E}|x|+\pi^2}{8|x|} \ , \\ 
&{\cal F} (\ln^4 |z|)=\bigg(\gamma_E^4+\frac{\gamma_E^2\pi^2}{2}+8\gamma_E\zeta_3+\frac{19\pi^4}{240}\bigg)\delta(x)+{\cal P}\frac{4\ln^3 e^{\gamma_E}|x|+\pi^2\ln e^{\gamma_E} |x|+8\zeta_3}{2|x|} \ ,
\end{align}
and so on. Clearly, the method above can be generalized to Fourier transform of ${\rm sign}(z)\ln^n |z|$ with minor modifications. In fact, the only formula one needs is 
\begin{align}
\int_{0}^{\infty} dz z^{\alpha-1} \cos zx=\frac{\Gamma(\alpha)\cos \frac{\pi \alpha}{2}}{|x|^{\alpha}} \ ,
\end{align}
when $0<\alpha<1$, which leads to
\begin{align}\label{eq:FTodd}
\int_{-\infty}^{\infty} {\rm sign}(z)|z|^{\alpha} \varphi(z)dz=2i\Gamma(\alpha)\cos \frac{\pi \alpha}{2}\int_{-\infty}^{\infty} dx \frac{\hat \varphi'(x)}{|x|^{\alpha}} \ ,
\end{align}
and one then proceeds exactly the same way by partial integrating. The results up to ${\rm sign(z)}\ln^2 |z|$ reads
\begin{align}
&{\cal F}(i\pi {\rm sign}(z))={\cal P}\frac{1}{x} \ , \\
&{\cal F}(i\pi{\rm sign}(z)\ln |z|)=-{\cal P}\frac{\ln e^{\gamma_E} |x|}{x} \ , \\ 
&{\cal F}(i\pi {\rm sign}(z)\ln^2 |z|)={\cal P}\frac{\ln^2e^{\gamma_E}|x|-\pi^2}{12x} \ .
\end{align}
Notice the absence of any subtraction term at $x=0$, which is expected for old distributions. 

\section{Converting the threshold ``plus function'' to principal value and numerical extraction of $c_H$}\label{sec:convert}
Like many other quantities in QFT, the quasi-PDF in massless perturbation theory has a high level of regularity in coordinate space: it is not only a tempered distribution\footnote{This is the level of regularity for a generic correlator in local quantum field theory~\cite{Streater:1989vi}.}, but also an analytic and local-integrable function with logarithmic small and large $z$ asymptotics. In particular, the large $z$ asymptotics for non-singlet quark quasi-distributions has the form $e^{-iP^z z}\ln^n z$ and corresponds exactly to the threshold limit. As a result,  when transformed to the momentum space in the sense of tempered distribution, the level of singularity at $x=1$ is enhanced, requiring additional subtraction terms. The spirit of such subtraction is the same as Fourier transform of logarithms and is normally presented in the ``plus-function'' notation in literature~\cite{Li:2020xml}. Since in the threshold limit the natural prescription is perhaps the principal value, which is adopted in this paper, we need to convert the ``plus-functions'' into principal value, defined in Eq.~(\ref{eq: defprin}). This is crucial to extract the coefficient of $\delta(1-y)$ term. In this appendix we present the details of this conversion. 

We first consider the region $y>1$. A plus function of the following form
\begin{align}
\bigg \langle \bigg[\theta(y-1)f(y)\bigg]^{[\infty]}_{\oplus (1)},\varphi(y) \bigg \rangle=\int_{1}^{\infty}f(y)\bigg(\varphi(y)-\varphi(1)\bigg) dy \ ,
\end{align}
can be converted to principal value in the threshold limit in the following way. First, one expand $f(y)$ at $y=1$ and separate the ${\cal O}(1-y)^{-1}$ singular term and non-singular term
\begin{align}
f(y)=f_{\rm sing}(y)+\bigg(f(y)-f_{\rm sing}(y)\bigg) \ ,
\end{align}
the first term survives the threshold limit in the region $1<y<2$, while the second term will only contribute to threshold limit through the $\delta(1-y)$ term, namely
\begin{align}
&\int_{1}^{\infty}f(y)\bigg(\varphi(y)-\varphi(1)\bigg) dy=\int_{1}^{2}f_{\rm sing}(y)\bigg(\varphi(y)-\varphi(1)\bigg)dy+ 
 \int_{2}^{\infty}f_{\rm sing}(y)\varphi(y)dy \nonumber \\ 
&-\bigg[\int_{1}^{2}\bigg(f(y)-f_{\rm sing}(y)\bigg)dy+\int_{2}^{\infty} f(y) dy\bigg]\varphi(1)+\int_{1}^{\infty}\bigg(f(y)-f_{\rm sing}(y)\bigg)\varphi(y) dy \ , 
\end{align}
clearly, the first line is just our convention of the principal value, while the last terms is power-suppressed. As a result, one has in the threshold limit
\begin{align}
&\bigg[\theta(y-1)f(y)\bigg]^{[\infty]}_{\oplus (1)}\nonumber \\ 
&\rightarrow {\cal P}\bigg(\theta(y-1)f_{\rm sing}(y)\bigg)-\bigg[\int_{1}^{2}\bigg(f(y)-f_{\rm sing}(y)\bigg)dy+\int_{2}^{\infty} f(y) dy\bigg]\delta(y-1) \ .
\end{align}
Similarly one has
\begin{align}
&\bigg[\theta(1-y)\theta(y)f(y)\bigg]^{[-\infty]}_{+(1)}\nonumber \\ 
&\rightarrow {\cal P}\bigg(\theta(1-y)\theta(y)f_{\rm sing}(y)\bigg)-\bigg[\int_{0}^{1}\bigg(f(y)-f_{\rm sing}(y)\bigg)dy\bigg]\delta(y-1) \ ,
\end{align}
and
\begin{align}
&\bigg[\theta(-y)\theta(y+1)f(y)\bigg]^{[-\infty]}_{+(1)}\rightarrow -\int_{-1}^{0}f(y)dy \delta(1-y) \ ,  \\ 
&\bigg[\theta(-y-1)f(y)\bigg]^{[-\infty]}_{+(1)}\rightarrow -\int_{-\infty}^{-1}f(y)dy \delta(1-y)  \  .
\end{align}
Given all above, it is not hard to convert all the threshold limit of quasi-PDF into principal values. All the terms depending on $L_z$ are easy, so we present here only the constant terms in $\delta(1-y)$.
\subsection{$C_AC_F$ term}
For this term, numerically one has
\begin{align}
&2(c_H+c_1)_{C_FC_A}+\frac{-528 \zeta_3 -3\pi ^4+100 \pi ^2}{216 \pi ^2}=0.6589 \ , \\ 
&c_H|_{C_FC_A}=-0.0840 \ ,
\end{align}
consistent with the numerical result in the main text. 
\subsection{$C_F^2$ term}
For this term, numerically one has
\begin{align}
&\frac{-5\pi^2+3\pi^4-120\zeta_3}{30\pi^2}+2c_H|_{C_F^2}=0.4782 \ , \\
&c_H|_{C_F^2}=0.0725 \ .
\end{align}
This is in agreement with the result in the main text.
\subsection{$C_Fn_fT_f$ term}
For this term, it is easy to check that
\begin{align}
l_0=\frac{-4\pi^2+24\zeta_3}{27\pi^2} +2(c_1+c_H)_{C_Fn_fT_F}=\frac{\zeta_3-\frac{13}{9}}{\pi ^2}-\frac{5}{54} \ ,
\end{align}
from which one has
\begin{align}
c_H|_{C_Fn_fT_F}=\frac{36 \zeta_3+51 \pi ^2+1312}{1296 \pi ^2}= 0.1453 \ ,
\end{align}
consistent with the numerical result in the main text.  In Appendix. \ref{sec:cHanaly} we extracts the $C_FC_A$ and the $C_F^2$ terms in $c_H$ analytically through the coordinate space representation.

\section{Results for the NNLO threshold limit in momentum space}\label{sec: NNLOmomemtum}
In this appendix we present the detailed results fore the quark non-singlet quasi-PDF 
in the threshold limit. Given the coordinate space expressions in Sec. \ref{sec:thresholcord} and the explicit form of $c_a$ in Eq.~(\ref{eq:resultca}), by Fourier transforming all the results to momentum space using the rules in Appendix \ref{sec:fourier} and \ref{sec:convert},  one obtains the threshold limit of the quasi-PDF\footnote{The threshold limit is independent of the spin structure, since both the heavy-light Sudakov form-factor and the jet function are independent of the spin structure.}
\begin{align}
&2\tilde f^{(2)}(y,L_z)\nonumber \\ 
&=\sum_{i=0}^{2}\bigg[C_FC_A\bigg({\cal P}g_i(y)+g_i\delta_y\bigg)+C_F^2\bigg({\cal P}h_i(y)+h_i\delta_y\bigg)+C_Fn_fT_F\bigg({\cal P}l_i(y)+l_i\delta_y\bigg)\bigg](-L_z)^i\ . 
\end{align}
Here the $L_z^{(0)}$ coefficients reads
\begin{align}
&g_0(y)|_{y>1}=\frac{-33 \ln ^2(y-1)+(100-3 \pi ^2)\ln (y-1)-27 \zeta_3+7 \pi ^2-12}{18 \pi ^2 (y-1)} \ , \\ 
&g_0(y)|_{y<1}=\frac{198 \ln ^2(1-y)+(18 \pi ^2-600) \ln (1-y)-432 \zeta_3+57 \pi ^2+1022}{108 \pi ^2 (y-1)} \ , \\ 
&h_0(y)|_{y>1}=\frac{24 \ln^3(y-1)-24 \ln ^2(y-1)+8 \pi ^2 \ln(y-1)+84 \zeta_3-11 \pi ^2+6}{12 \pi ^2 (y-1)} \ , \\
&h_0(y)|_{y<1}=\frac{-24 \ln ^3(1-y)+48 \ln ^2(1-y)-(8 \pi ^2 +24)\ln(1-y)-12 \zeta_3-3 \pi ^2+6}{12 \pi ^2 (y-1)}\ , \\
&l_0(y)|_{y>1}=\frac{6 \ln^2(y-1)-16 \ln(y-1)-\pi ^2+3}{9 \pi ^2 (y-1)} \ , \\ 
&l_{0}(y)|_{y<1}=\frac{-18 \ln^2(1-y)+48 \ln (1-y)-6 \pi ^2-85}{27 \pi ^2 (y-1)}\ .
\end{align}
For the $-L_z$ term one has
\begin{align}
&g_{1}(y)|_{y>1}=\frac{11 \ln(y-1)}{6 \pi ^2 (y-1)} \ , \\ 
&g_1(y)|_{y<1}=\frac{-33 \ln (1-y)-3 \pi ^2+100}{18 \pi ^2 (y-1)}\ , \\ 
&h_1(y)|_{y>1}=\frac{-6 \ln ^2(y-1)-\pi ^2}{3 \pi ^2 (y-1)} \ , \\
&h_{1}(y)|_{y<1}=\frac{12 \ln ^2(1-y)-12 \ln(1-y)+\pi ^2}{3 \pi ^2 (y-1)} \ , \\
&l_{1}(y)|_{y>1}=-\frac{2 \ln(y-1)}{3 \pi ^2 (y-1)}\ , \\ 
&l_{1}(y)|_{y<1}=\frac{6 \ln (1-y)-16}{9 \pi ^2 (y-1)}\ .
\end{align}
Finally, for the $L_z^2$ term one has
\begin{align}
&g_{2}(y)|_{y<1}=-\frac{11}{12\pi^2(1-y)} \ , \\ 
&h_{2}(y)|_{y<1}=\frac{2\ln(1-y)}{\pi^2(1-y)} \ , \\
&l_{2}(y)|_{y<1}=\frac{1}{3\pi^2(1-y)} \ .
\end{align}
Notice that $L_z^2$ term only appear in the ``physical region''. Finally, for the $\delta_y \equiv \delta(1-y)$ term, one has
\begin{align}
&g_0+g_1L_z=\frac{\left(-18 \zeta_3+\pi ^2-8\right)}{12 \pi ^2}L_z+\frac{-528 \zeta_3 -3\pi ^4+100 \pi ^2}{216 \pi ^2}+2(c_1+c_H)_{C_FC_A} \ , \\
&h_0+h_1L_z+h_2L_z^2=-\frac{L_z^2}{6}+\frac{2-\pi^2+28\zeta_3}{4\pi^2}L_z+\frac{-5\pi^2+3\pi^4-120\zeta_3}{30\pi^2}+2c_H|_{C_F^2}  \ , \\ 
&l_0+l_1L_z=\frac{1}{3\pi^2}L_z+\frac{-4\pi^2+24\zeta_3}{27\pi^2} +2(c_1+c_H)_{C_Fn_fT_F}\ .
\end{align}
The above provides a complete presentation of the threshold limit predicted by our factorization formalism and matches precisely with the two-loop exact result.

\section{Analytical extraction of $c_H$}\label{sec:cHanaly}
Although the momentum-space representation can be used to determine all the $L_z$ and $L_z^2$ terms in $\delta(1-y)$ analytically, it is relatively hard to obtain from that representation the constant terms $c_H$ in $\delta(1-y)$ without numerical integration, except for the $C_Fn_fT_F$ term. In this appendix we extract the remaining part of $c_H$ in Eq.~(\ref{hardtwoloop}) analytically using the coordinate space representation. 

The starting point is the coordinate space representation in Ref.~\cite{Li:2020xml}, for example for the $C_AC_F$ term one has\footnote{The $a_{020}^{(2)}$ term is trivial therefore omitted in the equation but will be added back at the end.}
\begin{align}
2\tilde f(z,L_z)|_{C_AC_F}^{(2)}=\sum_{i=0}^2l_z^i\int_{0}^1 g_i^a(t)\left(e^{-i\lambda t}-e^{-i\lambda}\right)dt+\sum_{i=0}^1l_z^i\int_{-1}^0 g_i^b(t)\left(e^{-i\lambda t}-e^{-i\lambda}\right)dt \ .
\end{align}
One needs to extract the $\lambda \rightarrow \infty$ limits in all integrals, corresponding to the threshold limit. For the purpose of extracting $c_H$, only $i=0$ term or $g^{a}_0\equiv g_a$, $g^{b}_0 \equiv g_b$ terms are needed. For $g_a$, as usual, one has
\begin{align}
&\int_{0}^1 g_a(t)\left(e^{-i\lambda t}-e^{-i\lambda}\right)dt\nonumber \\ 
&\rightarrow \int_{0}^1 g_{a,{\rm sing}}(t)\left(e^{-i\lambda t}-e^{-i\lambda}\right)dt-e^{-i\lambda}\int_{0}^1 \bigg(g_{a}-g_{a,{\rm sing}}\bigg)(t)dt \ .
\end{align}
The first term will contributes to $e^{-i\lambda}\ln^n \lambda$ in the  threshold limit and is relatively easy. The second term is relatively hard. We first consider the first term, to compare with our coordinate result one needs
\begin{align}
\ln (\pm i\lambda)=\frac{1}{2}L_z+\frac{l_z}{2} \pm \frac{i\pi}{2}{\rm sign}(z)-\gamma_E \ ,
\end{align}
which implies that for $L_z=0$
\begin{align}
&\int_{0}^{1} dt \frac{e^{-it\lambda}-e^{-i\lambda}}{1-t}\rightarrow e^{-i\lambda} \left(\frac{l_z}{2}+\frac{i\pi}{2}{\rm sign}(z)\right) \ , \\
&\int_{0}^{1} dt \frac{e^{-it\lambda}-e^{-i\lambda}}{1-t}\ln(1-t)\rightarrow -e^{-i\lambda} \frac{\pi^2}{24}+{\cal O}(l_z) \ , \\
&\int_{0}^{1} dt \frac{e^{-it\lambda}-e^{-i\lambda}}{1-t}\ln(1-t)\rightarrow -e^{-i\lambda} \frac{2\zeta_3}{3}+{\cal O}(l_z,{\rm sign}(z)) \ .
\end{align}
As a result, using the explicit form of the singular part of $g_a$, it contributes to
\begin{align}
\int_{0}^1 g_{a,{\rm sing}}(t)\left(e^{-i\lambda t}-e^{-i\lambda}\right)dt \rightarrow \bigg(\frac{22 \zeta_3}{9 \pi ^2}+\frac{600-18 \pi ^2}{1296}\bigg)e^{-i\lambda} \ ,
\end{align}
for the pure constant term (which means $l_z=L_z={\rm sign}(z)=0$ after using ${\rm sign}^2(z)=1$).
Similarly, for $g_b$ one has
\begin{align}
\int_{-1}^0 g_b(t)\left(e^{-i\lambda t}-e^{-i\lambda}\right)dt\rightarrow -e^{-i\lambda }\int_{-1}^{0} dt g_b(t) \ .
\end{align}
One can check that most of the integrals can be performed. The difficult integrals are
\begin{align}
&I_1=\int_{0}^1\frac{{\rm Li}_3(-t)-{\rm Li}_3(-1)}{1-t}dt=\int_{0}^{1}\frac{\ln(1-t)}{t}{\rm Li}_2(-t) dt \ , \\
&I_2=\int_{0}^1{\rm Li}_2(1-t)\frac{\ln(1+t)}{t}dt \  .
\end{align}
In fact, these integrals can also be performed in terms of ${\rm Li}_4\left(\frac{1}{2}\right)$. The crucial thing is the explicit formulas for two alternating Euler sums~\cite{valean_nahin_2019}
\begin{align}
&\sum_{n=1}^{\infty}(-1)^{n-1}\frac{H_n}{n^3}=\frac{11}{4}\zeta_4-\frac{7\ln 2}{4} \zeta_3+\frac{\ln^2 2}{2}\zeta_2-\frac{\ln^4 2}{12} -2{\rm Li}_4\left(\frac{1}{2}\right) \ , \\
&\sum_{n=1}^{\infty}(-1)^{n-1}\frac{H_n^{(2)}}{n^2}=-\frac{51}{16}\zeta_4+\frac{7\ln 2}{2} \zeta_3-\ln^2 2 \zeta_2+\frac{\ln^4 2}{6} +4{\rm Li}_4\left(\frac{1}{2}\right) \ .
\end{align}
Given these, all the difficult integrals above can be expressed in terms of special values of zeta-function up to ${\rm Li}_4\left(\frac{1}{2}\right)$.
\subsection{$C_FC_A$ term}
We first consider the $0<t<1$ region, namely, the $g_a(t)$ term. We split the terms into three categories $g_a=g_{a1}+g_{a2}+g_{a3}$: $g_{a1}$ includes all the algebraic and logarithmic terms, $g_{a2}$ includes all the ${\rm Li}_2$ and $g_{a3}$ includes all the ${\rm Li}_3$. The first categories can be even spilt to three subcategories as $g_{a1}=g_{a11}+g_{a12}+g_{a13}$: $g_{a11}$ contains all the algebraic terms and powers of $\ln(1-t)$, $g_{a12}$ contains all the powers of $\ln t$,  and $g_{a13}$ contains all the products of logarithms with two different arguments such as $\ln t \ln(1-t)$. One has
\begin{align}
\pi^2 g_{a11}=&\frac{\left(7 t^2+3\right) \zeta_3}{4 (t-1)}+\frac{\left(t^2+1\right) \left(-\frac{1}{3} \pi ^2 \ln (1-t)-2 \zeta_3\right)}{2 (t-1)}-\frac{3 \left(t^2+1\right) \left(\frac{1}{6} \pi ^2 \ln (1-t)+\zeta_3\right)}{2 (t-1)}\nonumber \\
&+\frac{\pi ^2 \left(1-2 t^2\right)}{24 (t-1)}-\frac{77 t^2-1104 t+77}{108 (t-1)}+\frac{\left(-3 t^2+6 t+85\right) \ln ^2(1-t)}{24 (t-1)}+\frac{\pi ^2 t^2 \ln (1-t)}{3(t-1)}\nonumber \\ 
&+\frac{\left(27 t^2-36 t+409\right) \ln (1-t)}{36 (t-1)} \ , \\ 
\pi^2g_{a12}=&\frac{\left(t^2+1\right) \ln ^3t}{24 (t-1)}+\frac{\left(17 t^2+24 t+11\right) \ln ^2t}{48 (t-1)}+\frac{\left(101 t^2+3 t+29\right) \ln t}{36 (t-1)} \ , \\ 
\pi^2 g_{a13}=&\frac{\left(t^2+1\right) \ln t \ln^2(1-t)}{2 (t-1)}-\frac{\left(3 t^2+6 t-53\right) \ln t \ln (1-t)}{12 (t-1)}+\frac{\ln^2 t \ln (1-t)}{t-1} \nonumber \\ 
&-(t+1) \ln t \ln (t+1) \ .
\end{align}
Only the first sub-category requires subtraction. Each sub-category can be evaluated and the total result reads
\begin{align}
&\int_{0}^1 dt \bigg(g_{a11}(t)-g_{a11,{\rm sing}}(t)+g_{a12}(t)+g_{a13}(t)\bigg)=\frac{-\frac{131 \zeta_3}{24}-\frac{11}{8}+2\ln 2}{\pi ^2}+\frac{\pi ^2}{30}+\frac{241}{216} \ .
\end{align}
We then move to term with ${\rm Li}_3$. The expression reads
\begin{align}
g_{a3}(t)=\frac{5(t^2+1) \text{Li}_3(1-t)}{2(t-1)\pi ^2}+\frac{2(t^2-1+2)) \text{Li}_3(-t)}{(t-1)\pi ^2}+\frac{\left(t+1\right) \text{Li}_3(t)}{\pi ^2} \ .
\end{align}
Notice the $t^2+1$ in the second term has been split as $t^2+1=t^2-1+2$. All the terms expect the $+2$ can be evaluated explicitly. The total result reads
\begin{align}
    &\int_{0}^1dt\bigg( g_{a3}(t)-g_{a3,{\rm sing}}(t) \bigg) dt=\nonumber \\ 
    &\frac{1}{288} \left(\frac{9 (96 \zeta_3+243-64 \ln4) }{\pi ^2}-16 \pi ^2-210\right)-\frac{4}{\pi^2}\int_0^1 \frac{\text{Li}_3(-t)-\text{Li}_3(-1)}{(1-t)} \, dt \ .
\end{align}
Notice the appearance of $I_1$ from the $2$. We then move to ${\rm Li}_2$ terms. The expression reads
\begin{align}
g_{a2}(t)=&\frac{ -\bigg((t^2-1)\ln t+2\ln t+(t^2-1)\bigg)\text{Li}_2(-t)}{\pi ^2(t-1)}\nonumber \\ 
&+\frac{ \bigg(\frac{53}{6}-t^2+(t^2-1)\ln(1-t)+2\ln(1-t)+(1-t^2)\ln t+2\ln t\bigg)\text{Li}_2(t)}{2\pi ^2(t-1)} \ .
\end{align}
Again, all terms can be explicitly integrated, except for the two $2\ln t$ terms. The total result can be expressed, after partial-integrating using the relations
\begin{align}
\frac{\ln t}{t-1}=-\frac{d}{dt}{\rm Li}_2(1-t) \ , \  \frac{d}{dt}\bigg(\text{Li}_2(t)-2 \text{Li}_2(-t)\bigg)=\frac{2 \ln (t+1)}{t}-\frac{\ln (1-t)}{t} \ ,
\end{align}
in terms of $I_2$ as
\begin{align}
&\int_{0}^{1} \bigg(g_{a2}(t)-g_{a2,{\rm sign}}(t)\bigg) dt = \nonumber \\ 
&\frac{1}{240} \left(\frac{5 (304 \zeta_3-33+96 \ln2)}{\pi ^2}-6 \pi ^2+5\right)+\frac{2}{\pi^2}\int_0^1 \frac{ \ln (1+t)}{t} \text{Li}_2(1-t)\, dt \ .
\end{align}
Notice the appearance of $I_2$. 

Similarly, for the $g_b$ integral in the region $-1<t<0$, using the same method one has the final result
\begin{align}
&\int_{-1}^{0} g_{b}(t)dt\nonumber \\ 
&=\frac{1}{960} \left(-\frac{5 \left(768 \text{Li}_4\left(\frac{1}{2}\right)+24 \zeta_3(19+12\ln2)+39+32 \ln^4(2)\right)}{\pi ^2}+34 \pi ^2+160 \left(2+\ln ^2(2)\right)\right) \nonumber \\ 
&+\frac{1}{\pi^2}\int_{-1}^0 \frac{\text{Li}_2(-t) \ln (-t)}{ 1-t} \, dt-\frac{2}{\pi^2}\int_{-1}^0 \frac{ \text{Li}_3(-t)}{ 1-t} \, dt \ .
\end{align}
Now,  after a few more partial integration, all the remaining integrals can be expressed in terms of
\begin{align}
&I_a=\frac{1}{\pi^2}\int_{0}^1 \frac{\ln(1+t)}{t}{\rm Li}_2(t) dt \ , \\ 
&I_b=\frac{1}{\pi^2}\int_{0}^1\frac{\ln(1+t)}{t}{\rm Li}_2(1-t) dt \ .
\end{align}
Expanding the logarithm, in terms of Harmonic numbers $H_n=\sum_{k=1}^n\frac{1}{k}$ and $H_n^{(2)}=\sum_{k=1}^{n}\frac{1}{k^2}$,  one finally has
\begin{align}
&I_a=\frac{\pi^2}{72}-\frac{1}{\pi^2}\sum_{n=1}^{\infty}\frac{(-1)^{n-1}H_n}{n^3}= \frac{2 \text{Li}_4\left(\frac{1}{2}\right)+\frac{7}{4} \zeta_3\ln (2)-\frac{\pi ^4}{60}+\frac{\ln ^4(2)}{12}-\frac{1}{12} \pi ^2 \ln ^2(2)}{\pi^2}\ , \\ 
&I_b=\frac{\pi^2}{72}-\frac{1}{\pi^2}\sum_{n=1}^{\infty}\frac{(-1)^{n-1}H_n^{(2)}}{n^2}=\frac{-4 \text{Li}_4\left(\frac{1}{2}\right)-\frac{7}{2} \zeta_3\ln (2)+\frac{71 \pi ^4}{1440}-\frac{1}{6} \ln ^4(2)+\frac{1}{6} \pi ^2 \ln ^2(2)}{\pi ^2} \ .
\end{align}
Combining all above, one finds that
\begin{align}
\int_{0}^1\bigg(g_a(t)-g_{a,{\rm sign}}(t)\bigg) dt&=\frac{1860 \zeta_3-34 \pi ^4+2655}{480 \pi ^2}+\frac{11}{27} \ , \\
\int_{-1}^{0} g_b(t) dt&=-\frac{456 \zeta_3-64 \pi ^2+39}{192 \pi ^2} \ , \\ 
\int_{0}^1\bigg(g_a(t)-g_{a,{\rm sign}}(t)\bigg) dt+\int_{-1}^{0} g_b(t) dt&=\frac{96 \zeta_3+341}{64 \pi ^2}+\frac{20}{27}-\frac{17 \pi ^2}{240} \ .
\end{align}
As expected, all the ${\rm Li}_4\left(\frac{1}{2}\right)$ and powers of $\ln 2$ cancels.  It is easy to check the correctness of such result numerically. Combining all the above, one has the asymptotics
\begin{align}
&-\left(\frac{96 \zeta_3+341}{64 \pi ^2}+\frac{20}{27}-\frac{17 \pi ^2}{240}\right)+\bigg(\frac{22 \zeta_3}{9 \pi ^2}+\frac{600-18 \pi ^2}{1296}\bigg)+\frac{1}{\pi^2}\bigg(\zeta_3-\frac{5\pi^2}{24} + \frac{4877}{576}\bigg) \nonumber \\ 
&=2(c_H+c_1)|_{C_FC_A} \ ,
\end{align}
which leads to
\begin{align}
c_H|_{C_FC_A}=\frac{241 \zeta_3}{144 \pi ^2}+\frac{11 \pi ^2}{320}-\frac{559}{1728}-\frac{971}{324 \pi ^2}=-0.084043046 \ ,
\end{align}
in agreement with the numerical result.
\subsection{$C_F^2$ term}
Similarly, one starts with\footnote{Again, the $a_{010}^{(2)}$ term is trivial and will be added only at the end.}
\begin{align}
2\tilde f(z,L_z)|_{C_F^2}^{(2)}=\sum_{i=0}^2l_z^i\int_{0}^1 h_i^a(t)\left(e^{-i\lambda t}-e^{-i\lambda}\right)dt+\sum_{i=0}^1l_z^i\int_{-1}^0 h_i^b(t)\left(e^{-i\lambda t}-e^{-i\lambda}\right)dt \ ,
\end{align}
and denote $h_{0}^a\equiv h_a$, $h_{0}^b \equiv h_b$. Using similar methods, one can show that for the $h_a$ term (when all the ${\rm Li}_2$ are transformed to be with arguments $z$ and $-z$, but not $1-z$),
\begin{align}
\int_{0}^1 \bigg(h_{a1}(t)-h_{a1,{\rm sign}}(t)\bigg) dt=-\frac{13 \zeta_3+2 \ln 4}{\pi ^2}-\frac{293}{144}-\frac{\pi ^2}{120}+\frac{3529}{192 \pi ^2} \ .
\end{align}
And for the ${\rm Li}_2$ terms
\begin{align}
&\int_{0}^1 \bigg(h_{a2}(t)-h_{a2,{\rm sign}}(t)\bigg) dt=\frac{1}{240} \left(-\frac{15 (8 \zeta_3+147+32 \ln 4)}{\pi ^2}+2 \pi ^2+145\right) \nonumber \\ 
&-\frac{4}{\pi^2}\int_{0}^1 \frac{\ln(1+t)}{t}{\rm Li}_2(1-t) dt \ .
\end{align}
The only difficult integral is $I_b$. Now, for the ${\rm Li}_3$ term one finally has
\begin{align}
&\int_{0}^1 \bigg(h_{a3}(t)-h_{a3,{\rm sign}}(t)\bigg) dt=\frac{1}{720} \left(\frac{45 (72 \zeta_3-95+64 \ln 4)}{\pi ^2}+94 \pi ^2-90\right) \nonumber \\ 
&+\frac{8}{\pi^2}\int_0^1 \frac{\text{Li}_3(-t)-\text{Li}_3(-1)}{(1-t)} \, dt \ .
\end{align}
Again, notice the appearance of $I_1$. Using the results for these integrals in the $C_AC_F$ section, one finally has
\begin{align}
\int_{0}^1 \bigg(h_{a}(t)-h_{a,{\rm sign}}(t)\bigg) dt=\frac{\frac{625}{192}-9 \zeta_3}{\pi ^2}+\frac{8 \pi ^2}{45}-\frac{14}{9} \ .
\end{align}
Again, all the $\ln 2$ and ${\rm Li}_4 \left(\frac{1}{2}\right)$ cancels. It is not hard to check this result numerically. For $h_{b}$, since one has $h_{b}=-2g_b$, one can use the result before
\begin{align}
\int_{-1}^0 dt h_b(t) dt=\frac{456 \zeta_3-64 \pi ^2+39}{96 \pi ^2} \ ,
\end{align}
which leads to
\begin{align}
\int_{0}^1 \bigg(h_{a}(t)-h_{a,{\rm sign}}(t)\bigg) dt+\int_{-1}^0 dt h_b(t) dt =\frac{703-816 \zeta_3}{192 \pi ^2}+\frac{8 \pi ^2}{45}-\frac{20}{9} \ .
\end{align}
This finishes the most difficult step. To extract the $c_H$, one still needs to obtain the large $\lambda$ asymptotics as the case with $C_{A}C_F$. For this purpose one needs one more integral
\begin{align}
\int_{0}^1 dt \frac{e^{-it\lambda}-e^{-i\lambda}}{1-t}\ln^3(1-t)\rightarrow -e^{-i\lambda}\frac{3\pi^4}{320}+{\cal O}(l_z,{\rm sign}(z)) \ .
\end{align}
Given these, one has
\begin{align}
2c_H|_{C_F^2}+\frac{\pi^2-36}{144}=-\frac{540 \zeta_3+23 \pi ^4-324 \pi ^2+360}{144 \pi ^2} \ ,
\end{align}
which leads finally to 
\begin{align}
c_H|_{C_F^2}=\frac{-45 \zeta_3-2 \pi ^4+30 \pi ^2-30}{24 \pi ^2}=0.0725181 \ ,
\end{align}
consistent with the numerical result. As a result, all the three terms for $c_H$ are determined analytically now:
\begin{align}
c_H=&\bigg(\frac{241 \zeta_3}{144 \pi ^2}+\frac{11 \pi ^2}{320}-\frac{559}{1728}-\frac{971}{324 \pi ^2}\bigg)C_FC_A+\bigg(\frac{-45 \zeta_3-2 \pi ^4+30 \pi ^2-30}{24 \pi ^2}\bigg)C_F^2 \nonumber \\ 
&+\bigg(\frac{36 \zeta_3+51 \pi ^2+1312}{1296 \pi ^2}\bigg)C_Fn_fT_F \ .
\end{align}
Combining with $c_a$, the above completely determines the heavy-light Sudakov kernel at two-loop, an universal object that also appears in TMD factorization for quasi-TMDPDFs and quasi-LFWFs.

\section{A sample two-loop diagram for the jet function}\label{sec:twoloopsample}
\begin{figure}[h!]
    \centering
    \includegraphics[height=6cm]{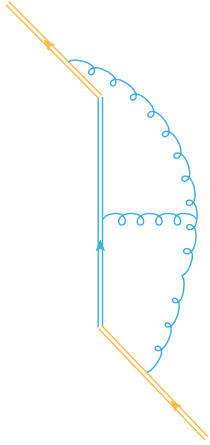}
    \caption{One of the two-loop diagrams for time like jet function, with a tripple gluon vertex}
    \label{fig:origin1}
\end{figure}
\begin{figure}
    \centering
    \includegraphics[height=6cm]{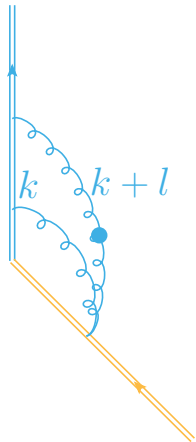}
    \caption{A  master integral from IBP reduction of Fig.~\ref{fig:origin1} with five propagators}
    \label{fig:master1}
\end{figure}
In this appendix,  we present calculation details for a sample two-loop diagram with triple gluon vertex and with six propagators, shown in Fig.~\ref{fig:origin1}. The purpose of this appendix is three fold. First, it provides a check on the non-trivial fact claimed in Ref.~\cite{Jain:2008gb}, that after flipping the incoming light-like gauge link direction from pointing to $+\infty n$ to $-\infty n$, the two-loop diagrams remain the same up to scaless terms. Second, it verifies in an explicit way, the all order result that after flipping of the light-like gauge-link, the time-like jet function relates to its space-like version through the analytic continuation $t\rightarrow -i|z|$ without encountering any singularities. Finally, it serves as a pedagogical introduction to a sample two-loop calculations with gauge-link propagators and light-cone singularities, which is hard to find in the literature. 

The starting point is Fig.~\ref{fig:origin1} . Using the standard Feynman rule, it is easy to check that the in momentum space ($t$ is conjugating to $q^0$) this diagram is proportional to
\begin{align}
F(111;110;100)\equiv \int \frac{d^Dkd^Dl}{(2\pi)^{2D}}\frac{1}{D_lD_{k+l}D_k}\frac{1}{(l\cdot v+q^0)((k+l)\cdot v+q^0)(l\cdot n)} \ .
\end{align}
The IBP reduction of this integral, after using once $\partial_{k}\cdot k$ and once $\partial_l\cdot l$, reads
\begin{align}
F(111;110;100)=&\frac{1}{D-4}\bigg(F(100;100;100)F(002;001;000)-F(021;110;100)\bigg) \nonumber \\ 
-&\frac{1}{(D-4)^2}\bigg(F(012;020;100)-F(102;020;100)\bigg) \ .
\end{align}
The convention for integrals $F(n_1n_2n_3;m_1m_2m_3;l_1l_2l_3)$ can be reads from the expressions above and follows roughly the same pattern as Ref.~\cite{Jain:2008gb} without the pre-factors.
Notice that all the denominators are with $+i0$ prescriptions, corresponding to one incoming light-like gauge link and one out going. This Wilson-line configuration is in fact much simpler than the one with both outgoing light-like gauge links and can be directly analytically continued into our space-like jet function without any discontinuity.  

In principle, the power of the denominators can be reduced more, but we will not proceed further in the IBP reduction since the level of difficulty is already not very high. In particular, all the integrals with four propagators are easy, so we focus on one of the ``difficult" integrals, $F(021;110;100)$ with five propagators. More explicitly, in momentum space one has 
\begin{align}
I\equiv F(021;110;100)=\mu_0^{8-2D}\int\frac{d^Dkd^Dl}{(2\pi)^{2D}}\frac{1}{D_{l+k}^2D_k}\frac{1}{(l\cdot v+q)((k+l)\cdot v+q)l\cdot n} \ .
\end{align}
Diagrammatically, it can be represented as in Fig.~(\ref{fig:master1}). The dot in the gluon propagator simply denotes the doubling $D_{k+l} \rightarrow D_{k+l}^2=((k+l)^2+i0)^2$. This diagram can be most efficiently evaluated in position space. For this purpose, one needs the coordinate space version of the doubled propagator
\begin{align}
G^{(2)}(-x^2+i0)=\int \frac{d^Dk}{(2\pi)^D}\frac{-ie^{-ik\cdot x}}{D_k^2}=\frac{\Gamma(\frac{D}{2}-2)}{2^{4-D}(4\pi)^{\frac{D}{2}}}(-x^2+i0)^{2-\frac{D}{2}} \ .
\end{align}
Remember the coordinate space version of the single propagator reads
\begin{align}
G(-x^2+i0)=\frac{\Gamma(\frac{D}{2}-1)}{2^{2-D}(4\pi)^{\frac{D}{2}}}(-x^2+i0)^{1-\frac{D}{2}}=\int \frac{d^Dk}{(2\pi)^D}\frac{ie^{-ik\cdot x}}{D_k} \ .
\end{align}
Given this, one actually has 
\begin{align}
I=i\int_{0}^{\infty} d\lambda_3 \int_{0}^{\infty} d\lambda_1 \int_{0}^{\lambda_1} d\lambda_2 G^{(2)}(-\lambda_1^2-\sqrt{2}\lambda_1\lambda_3 +i0)G(-\lambda_2^2-\sqrt{2}\lambda_2\lambda_3 +i0)e^{i\lambda_1q^0} \ .
\end{align}
Clearly, for $q^0<0$, it is possible to simply analytic continue in $\lambda_i=-i\lambda_{i}$ to obtain
\begin{align}\label{eq:defI}
&I=-\mu_0^{8-2D}|q^0|^{2D-9}\frac{\Gamma(\frac{D}{2}-2)\Gamma(\frac{D}{2}-1)}{\sqrt{2}(4\pi)^D2^{6-2D}}\nonumber \\ 
& \times \int_{0}^{\infty} d\lambda_3 \int_{0}^{\infty} d\lambda_1 \int_{0}^{\lambda_1} d\lambda_2 (\lambda_1^2+\lambda_1\lambda_3)^{2-\frac{D}{2}}(\lambda_2^2 +\lambda_2\lambda_3)^{1-\frac{D}{2}}e^{-\lambda_1} \ .
\end{align}
Introducing the parameterization 
\begin{align}
\lambda_2=\lambda_1 x \ ,  \ 0<x<1 \ , \\ 
\lambda_3=x\lambda_1 \rho \ , \ 0<\rho<\infty \ ,
\end{align}
one obtains
\begin{align}
I=-\mu_0^{8-2D}|q^0|^{2D-9}\frac{\Gamma(\frac{D}{2}-2)\Gamma(\frac{D}{2}-1)\Gamma(9-2D)}{\sqrt{2}(4\pi)^D2^{6-2D}}\int_{0}^{\infty} d\rho\int_{0}^1 dx x^{3-D}(1+\rho)^{1-\frac{D}{2}}(1+x\rho)^{2-\frac{D}{2}} \ .
\end{align}
Now, replacing  $x$ by $x\rightarrow 1-y$ and re-introducing through $\rho=\frac{x}{1-x}$, one has
\begin{align}
&I=-\mu_0^{8-2D}|q^0|^{2D-9}\frac{\Gamma(\frac{D}{2}-2)\Gamma(\frac{D}{2}-1)\Gamma(9-2D)}{\sqrt{2}(4\pi)^D2^{6-2D}}M \ , \\
&M=\int_{0}^{1} dx \int_{0}^1dy (1-x)^{-1-2\epsilon}(1-y)^{-1+2\epsilon}(1-xy)^{\epsilon} \ .
\end{align}
To make further simplification, one would like to partial integrate with respect to $x$ by writing $(1-x)^{-1-2\epsilon}=\frac{1}{2\epsilon}\frac{d}{dx}(1-x)^{-2\epsilon}$. Then one obtains
\begin{align}
M=-\frac{1}{4\epsilon^2}-&\frac{1}{2}\int_{0}^1 dx\int_{0}^1 dy(1-x)^{-2\epsilon}(1-y)^{2\epsilon}(1-xy)^{-1+\epsilon}\nonumber \\ 
+&\frac{1}{2}\int_{0}^1 dx\int_{0}^1 dy(1-x)^{-2\epsilon}(1-y)^{-1+2\epsilon}(1-xy)^{-1+\epsilon} \ .
\end{align}
Notice that the rule of the dimensional regularization $0^{a+b\epsilon}=0$ and $1^{a+b\epsilon}=1$ has been used. To evaluate the simpler one among the two integrals, the
\begin{align}
 M_1=\int_{0}^1 dx\int_{0}^1 dy(1-x)^{-2\epsilon}(1-y)^{2\epsilon}(1-xy)^{-1+\epsilon} \ ,
\end{align}
simply notice that this integral is absolutely and uniformly convergent in the strip regions $-\frac{1}{2}+\delta<{\rm Re} (\epsilon)<\frac{1}{2}-\delta$ for any $0<\delta<\frac{1}{2}$, therefore it defines an analytic function around $\epsilon=0$ with all the Taylor coefficients obtainable through expanding before integrating. As a result one obtains
\begin{align}
 M_1=\frac{\pi^2}{6}-\zeta_3\epsilon+\int_{0}^1\int_{0}^{1} dxdy\frac{\left(2\ln(1-y)-2\ln(1-x)+\ln(1-xy)\right)^2}{2(1- x y)}\epsilon^2 +{\cal O}(\epsilon^3)\ .
\end{align}
The first term is due to 
\begin{align}
\int_{0}^{1}\int_{0}^1 dxdy\frac{1}{1-xy}=\zeta_2=\frac{\pi^2}{6} \ ,
\end{align}
while the second term is due to
\begin{align}
\int_{0}^{1}\int_{0}^1 dxdy\frac{2\ln(1-x)-2\ln(1-y)-\ln (1-xy)}{1-xy}=-\zeta_3 \ .
\end{align}
To evaluate the third term, it is instructive to list all the necessary pieces
\begin{align}
\int_{0}^1\int_{0}^{1} dxdy \frac{\ln^2 (1-x)}{1-xy}=\frac{\pi^4}{15} \ , \\ 
\int_{0}^{1}\int_{0}^{1} dxdy\frac{\ln^2(1-xy)}{1-xy}=\frac{\pi^4}{45} \ , \\
\int_{0}^{1}\int_{0}^{1}dxdy \frac{\ln(1-y)\ln(1-x)}{1-xy}=\frac{17\pi^4}{360} \ .
\end{align}
Given all above, one finally has
\begin{align}
M_1&=\frac{\pi^2}{6}-\zeta_3\epsilon +\pi^4\bigg(\frac{4}{15}+\frac{1}{90}-\frac{17}{90}\bigg)\epsilon^2 +{\cal O}(\epsilon^3) \nonumber \\ 
&=\frac{\pi^2}{6}-\zeta_3\epsilon+\frac{4\pi^4}{45}\epsilon^2+{\cal O}(\epsilon^3)
\end{align}
One then move to the more difficult integral
\begin{align}
M_2=\int_{0}^1 dx\int_{0}^1 dy(1-x)^{-2\epsilon}(1-y)^{-1+2\epsilon}(1-xy)^{-1+\epsilon} \ .
\end{align}
To evaluate this integral, it is more convenient to transform it into a single infinite sum. This can be achieved using
\begin{align}
(1-xy)^{-1+\epsilon} =-\sum_{n=0}^{\infty} \frac{\Gamma(\epsilon)}{\Gamma(-\epsilon)\Gamma(1+\epsilon)}\frac{\Gamma(1+n-\epsilon)}{\Gamma(n+1)}(xy)^{n} \ .
\end{align}
Integrating term by term, one get
\begin{align}
&M_2=\frac{\Gamma(1-2\epsilon)\Gamma(\epsilon)\Gamma(2\epsilon)\sin \pi \epsilon}{\pi}S(\epsilon) \ , \\ 
&S(\epsilon)=\sum_{n=0}^{\infty}\frac{\Gamma(1+n)}{\Gamma(1+n+2\epsilon)\Gamma(2+n-2\epsilon)}\Gamma(1+n-\epsilon) \ .
\end{align}
The sum converges absolutely for ${\rm Re}(\epsilon)>0$, but develops a pole at $\epsilon=0$. To extract the ${\cal O}(\epsilon^3)$ contribution one can first expand
\begin{align}
&\frac{\Gamma(1+n)}{\Gamma(1+n+2\epsilon)\Gamma(2+n-2\epsilon)}=\frac{1}{\Gamma(2+n)}\nonumber \\ 
&+\frac{2 \epsilon}{(n+1)^2 \Gamma (n+1)}+\frac{\epsilon^2 \left(4-4 (n+1)^2 \psi ^{(1)}(n+1)\right)}{(n+1)^3 \Gamma (n+1)}+\frac{\epsilon^3 \left(8-8 (n+1)^2 \psi ^{(1)}(n+1)\right)}{(n+1)^4 \Gamma (n+1)} \ .
\end{align}
Now, one can sum term by term, and in the third term one simply set $\epsilon=0$ in $\Gamma(1+n-\epsilon)$. The reason is that the decay of the polygamma function makes all the terms except the first one absolutely convergent at $\epsilon=0$.
Using the relations
\begin{align}
&\sum_{n=0}^{\infty}\frac{\Gamma(1+n-\epsilon)}{\Gamma(2+n)}=\frac{\Gamma(1-\epsilon)}{\epsilon} \ , \\ 
&\sum_{n=0}^{\infty}\frac{2\epsilon\Gamma(1+n-\epsilon)}{(n+1)\Gamma(n+2)}=2 \Gamma (1-\epsilon) \bigg(\psi(\epsilon+1)+\gamma_E \bigg) \ , \\ 
&\sum_{n=0}^{\infty}\frac{1- (n+1)^2 \psi ^{(1)}(n+1)}{(n+1)^3 \Gamma (n+1)}4\epsilon^2\Gamma(1+n-\epsilon)=\frac{2}{3} \Gamma (-\epsilon) \left(\pi ^2 \epsilon^2-6 \epsilon(\psi(\epsilon)+\gamma_E )-6\right) \ , \\
&\sum_{n=0}^{\infty}\frac{8-8 (n+1)^2 \psi ^{(1)}(n+1)}{(n+1)^4} =-\frac{\pi^4}{15} \ ,
\end{align}
one finally obtains
\begin{align}
S(\epsilon)=&\frac{1}{\epsilon}+\gamma_E+\frac{1}{12}\left(6\gamma_E^2+5\pi^2\right)\epsilon+\frac{-68 \zeta_3+2 \gamma_E ^3+5 \gamma_E  \pi ^2}{12}\epsilon^2\nonumber \\ 
&+\frac{-8160 \gamma_E  \zeta_3+60 \gamma_E ^4+49 \pi ^4+300 \gamma_E ^2 \pi ^2}{1440}\epsilon^3 \ .
\end{align}
Combining with the coefficients, one obtains
\begin{align}
M_2=\frac{1}{2\epsilon^2}+\frac{\pi^2}{2}-3\zeta_3\epsilon+\frac{4\pi^4}{15}\epsilon^2+{\cal O}(\epsilon^3) \ , 
\end{align}
which leads to
\begin{align}
M=-\frac{1}{4\epsilon^2}-\frac{M_1}{2}+\frac{M_2}{2}=\frac{\pi^2}{6}-\zeta_3\epsilon+\frac{4\pi^4}{45}\epsilon^2+{\cal O}(\epsilon^3) \ .
\end{align}
The above completely determines the momentum space version of the integral. 
In fact, converting to the notation in Ref.~\cite{Jain:2008gb}, using $a=2q^0$, one has
\begin{align}
&\hat F(021;110;100)=\mu^{4\epsilon}e^{\gamma_E\epsilon}(4\pi)^D \frac{q^0}{2}n\cdot v I \ , \\ 
&=-\mu^{4\epsilon}e^{\gamma_E\epsilon}(4\pi)^D \frac{q^0}{2}|q^0|^{2D-9}\frac{\Gamma(\frac{D}{2}-2)\Gamma(\frac{D}{2}-1)\Gamma(9-2D)}{2^{7-2D}(4\pi)^D}M \ , \\ 
&=\left(\frac{\mu}{2|q^0|}\right)^{4\epsilon}\frac{\Gamma(\frac{D}{2}-2)\Gamma(\frac{D}{2}-1)\Gamma(9-2D)}{2^{8-2D}}(16e^{\gamma_E})^{\epsilon}M \nonumber \\ 
&=\left(\frac{\mu}{-a}\right)^{4\epsilon}\left(-\frac{\pi^2}{6\epsilon}+\zeta_3-\frac{61\pi^4}{180}\epsilon+{\cal O}(\epsilon^2)\right) \ .
\end{align}
 The ${\cal O}(\epsilon)$ term is required in order to obtain the constant contribution to $\hat F(111;110;100)$ due to the $\frac{1}{4-D}$ prefactor.

One also notice that to determine the same diagram directly in coordinate space, one simply needs to Fourier transform back 
\begin{align}
I(t)=\int_{-\infty}^{\infty}  \frac{dq^0}{2\pi} e^{-iq^0t} I(q^0) \ ,
\end{align}
which results in 
\begin{align}
I(t)=i\int_{0}^{\infty} d\lambda_3 \int_{0}^{t} d\lambda_2 G^{(2)}(-t^2-\sqrt{2}t\lambda_3 +i0)G(-\lambda_2^2-\sqrt{2}\lambda_2\lambda_3 +i0) \ .
\end{align}
Now, the analytic continuation in $t\rightarrow -i|z|$ can be performed easily, which gives 
\begin{align}
iI(|z|)=\int_{0}^{\infty} d\lambda_3 \int_{0}^{|z|} d\lambda_2 G^{(2)}(z^2+\sqrt{2}z\lambda_3 +i0)G(\lambda_2^2+\sqrt{2}\lambda_2\lambda_3) \ ,
\end{align}
and
\begin{align}
&iI(|z|)=(\mu_0 |z|)^{8-2D}\frac{\Gamma(\frac{D}{2}-2)\Gamma(\frac{D}{2}-1)}{\sqrt{2}(4\pi)^D2^{6-2D}}\int_{0}^{\infty} d\rho\int_{0}^1 dx x^{3-D}(1+\rho)^{1-\frac{D}{2}}(1+x\rho)^{2-\frac{D}{2}} \nonumber \\
&=(\mu_0 |z|)^{8-2D}\frac{\Gamma(\frac{D}{2}-2)\Gamma(\frac{D}{2}-1)}{\sqrt{2}(4\pi)^D2^{6-2D}}\bigg(\frac{\pi^2}{6}-\zeta_3\epsilon+\frac{4\pi^4}{45}\epsilon^2+{\cal O}(\epsilon^3)\bigg) \ ,
\end{align}
which is again determined by the same integral. As a result, the coordinate space version is also completely determined. It is easy to check that the above is simply the same integral resulting from IBP reduction of our space-like jet function. 

Finally, we address the issue of the flipping the sign of $n$. Clearly, in case of $F(021;110;100)$ it is suffice to show that 
\begin{align}
 &\int_{0}^{\infty} d\lambda_3 (\lambda_1^2+\lambda_1\lambda_3)^{2-\frac{D}{2}}(\lambda_2^2 +\lambda_2\lambda_3)^{1-\frac{D}{2}}\nonumber \\ 
 &=-\int_{0}^{\infty} d\lambda_3 (\lambda_1^2-\lambda_1\lambda_3-i0)^{2-\frac{D}{2}}(\lambda_2^2 -\lambda_2\lambda_3-i0)^{1-\frac{D}{2}}\ .
\end{align}
After exponentiation, the difference is easily seen to be proportional to
\begin{align}
\int_{-\infty}^{\infty} d\lambda_3 e^{i\lambda_3\left(\alpha_1\lambda_1+\alpha_2\lambda_2\right)}=2\pi \delta(\alpha_1\lambda_1+\alpha_2\lambda_2) \ .
\end{align}
However, since $\alpha_1\lambda_1+\alpha_2\lambda_2 \ge 0$, the delta-function after integrating with the remaining parameters simply vanishes due to the rule $0^{a\epsilon}\equiv 0$ for any $a \ne 0$ in DR. It is not hard to check this for all the remaining terms.

\bibliographystyle{apsrev4-1}
\bibliography{bibliography.bib}

\end{document}